\documentclass[12pt]{article}

\usepackage{newtxtext,newtxmath}

\usepackage{graphicx}

\usepackage[letterpaper,margin=1in]{geometry}

\renewenvironment{abstract}
	{\quotation}
	{\endquotation}

\date{}

\makeatletter
\renewcommand{\fnum@figure}{\textbf{Figure \thefigure}}
\renewcommand{\fnum@table}{\textbf{Table \thetable}}
\makeatother

\usepackage{scicite}

\usepackage{url}

\def\arcsec{\hbox{$^{\prime\prime}$}}

\newcommand{\frb}{\rm FRB~20240304B}

\newcommand{\dmunits}{\ensuremath{{\rm pc \, cm^{-3}}}}
\newcommand{\dmcosmic}{\ensuremath{{\rm DM}_{\rm cosmic}{(z)}}}
\newcommand{\dmcosmicfrb}{\ensuremath{{\rm DM}_{\rm cosmic}{(z_{\rm FRB})}}}

\newcommand{\dmfrb}{\ensuremath{{\rm DM}_{\rm FRB}}}

\newcommand{\dmhost}{\ensuremath{{\rm DM}_{\rm host}}}

\newcommand{\dmmwism}{\ensuremath{{\rm DM}_{\rm MW,ISM}}}
\newcommand{\dmhalo}{\ensuremath{{\rm DM}_{\rm MW,halo}}}

\def\arcsec{\hbox{$^{\prime\prime}$}}

\newcommand\arcmin{\hbox{$.\!\!^{\prime}$}}

\newcommand{\sfunits}{\ensuremath{{\rm M}_{\odot}}\,{\rm yr^{-1}}}

\newcommand{\pox}{\ensuremath{P(O|x)}}
\newcommand{\pux}{\ensuremath{P(U|x)}}
\newcommand{\pu}{\ensuremath{P(U)}}
\newcommand{\puz}{\ensuremath{P(U|z)}}
\newcommand{\pzdm}{\ensuremath{p(z|\rm{DM})}}
\newcommand{\pathpost}{\ensuremath{97.5\%}} 
\newcommand{\puxpost}{\ensuremath{2.4\%}}

\newcommand{\rmunits}{\ensuremath{{\rm rad \, m^{-2}}}}

\newcommand{\redshift}{\ensuremath{z_{\mathrm{spec}}=2.148\pm0.001}}

\newcommand{\OVI}{[\hbox{{\rm O}\kern 0.1em{\sc vi}}]}

\newcommand{\NV}{\hbox{{\rm N}\kern 0.1em{\sc v}}}
\newcommand{\SiIV}{\hbox{{\rm Si}\kern 0.1em{\sc iv}}}
\newcommand{\OIV}{[\hbox{{\rm O}\kern 0.1em{\sc iv}}]}
\newcommand{\NIV}{[\hbox{{\rm N}\kern 0.1em{\sc iv}}]}
\newcommand{\CIV}{\hbox{{\rm C}\kern 0.1em{\sc iv}}}
\newcommand{\HeII}{\hbox{{\rm He}\kern 0.1em{\sc ii}\kern 0.1em{$\lambda1640$} }}
\newcommand{\OIII}{[\hbox{{\rm O}\kern 0.1em{\sc iii}}]{$\lambda5007$}}
\newcommand{\OIIId}{[\hbox{{\rm O}\kern 0.1em{\sc iii}}]{$\lambda4959\lambda5007$}}
\newcommand{\NIII}{[\hbox{{\rm N}\kern 0.1em{\sc iii}}]}
\newcommand{\AlIII}{\hbox{{\rm Al}\kern 0.1em{\sc iii}}}
\newcommand{\SiIII}{\hbox{{\rm Si}\kern 0.1em{\sc iii}}}
\newcommand{\CIII}{\hbox{{\rm C}\kern 0.1em{\sc iii}]}}
\newcommand{\NeIV}{[\hbox{{\rm Ne}\kern 0.1em{\sc iv}}]}
\newcommand{\MgII}{\hbox{{\rm Mg}\kern 0.1em{\sc ii}}}

\newcommand{\CII}{[\hbox{{\rm C}\kern 0.1em{\sc ii}]}}

\newcommand{\He}{\hbox{{\rm He}\kern 0.1em{\sc ii}\kern 0.1em{$\lambda1640\lambda4686$}}}

\newcommand{\msol}{M$_\odot$}

\newcommand{\zsol}{Z$_\odot$}

\newcommand{\fastpp}{{\tt FAST++}}
\newcommand{\prospector}{{\tt Prospector}}
\newcommand{\slinefit}{{\tt slinefit}}
\newcommand{\Halpha}{H$\alpha$}
\newcommand{\Hbeta}{H$\beta$}
\newcommand{\SII}{[\hbox{{\rm S}\kern 0.1em{\sc ii}}]$\lambda6716\lambda6731$}
\newcommand{\NII}{[\hbox{{\rm N}\kern 0.1em{\sc ii}}]}
\newcommand{\NIId}{[\hbox{{\rm N}\kern 0.1em{\sc ii}}]$\lambda6548\lambda6584$}
\newcommand{\OII}{[\hbox{{\rm O}\kern 0.1em{\sc ii}}]}
\newcommand{\MgI}{\hbox{{\rm Mg}\kern 0.1em{\sc i}}}
\newcommand{\FeII}{\hbox{{\rm Fe}\kern 0.1em{\sc ii}}}

\newcommand{\OI}{\hbox{{\rm O}\kern 0.1em{\sc i}}}
\newcommand{\NeII}{[\hbox{{\rm Ne}\kern 0.1em{\sc ii}}] }
\newcommand{\NaI}{[\hbox{{\rm Na}\kern 0.1em{\sc i}}] }
\newcommand{\NeIII}{[\hbox{{\rm Ne}\kern 0.1em{\sc iii}}] }

\def\scititle{
	A fast radio burst from the first 3 billion years of the Universe
}
\title{\bfseries \boldmath \scititle}

\author{
Manisha Caleb$^{1,2\ast}$, 
Themiya	Nanayakkara$^{3,4}$,
Benjamin Stappers$^5$, \and
In\'es Pastor-Marazuela$^5$,
Ilya S. Khrykin$^6$,
Karl Glazebrook$^{3}$, 
Nicolas	Tejos$^6$, \and
J. Xavier Prochaska$^{7,8,9}$,
Kaustubh Rajwade$^{10}$,
Lluis Mas-Ribas$^{7,11}$, \and
Laura N. Driessen$^1$, 
Wen-fai	Fong$^{12}$,
Alexa C. Gordon$^{13}$, 
Jordan Hoffmann$^{13}$, \and
Clancy W. James$^{13}$,
Fabian Jankowski$^{14}$, 
Lordrick Kahinga$^{7,15}$, \and
Michael	Kramer$^{16,5}$, 
Sunil Simha$^{12,16}$,
Ewan D. Barr$^{17}$,  \and
Mechiel Christiaan Bezuidenhout$^{18}$,
Xihan Deng$^{19}$,
Zeren Lin$^{19}$,  \and
Lachlan Marnoch$^{20,21,22,23}$, 
Christopher D. Martin$^{19}$,
Anya Nugent$^{24}$, \and
Kavya Shaji$^{1,21}$,
Jun Tian$^5$ \and
\small$^{1}$ Sydney Institute for Astronomy, School of Physics, The University of Sydney, Sydney, NSW 2006, Australia \and
\small$^{2}$ ARC Centre of Excellence for Gravitational Wave Discovery (OzGrav), Hawthorn, VIC 3122, Australia \and
\small$^{3}$ Centre for Astrophysics and Supercomputing, Swinburne University of Technology, Hawthorn, VIC 3122, Australia \and
\small$^{4}$ JWST Australian Data Centre (JADC), Swinburne Advanced Manufacturing and Design Centre (AMDC), \and
\small{John Street, Hawthorn, VIC 3122, Australia} \and
\small$^{5}$ Jodrell Bank Centre for Astrophysics, Department of Physics and Astronomy, The University of Manchester, \and 
\small{Oxford Road, Manchester M13 9PL, United Kingdom} \and
\small$^{6}$ Instituto de F\'isica, Pontificia Universidad Cat\'olica de Valpara\'iso, Casilla 4059, Valpara\'iso, Chile \and
\small$^{7}$ Department of Astronomy and Astrophysics, University of California, 1156 High Street, Santa Cruz, CA 95064, USA \and
\small$^{8}$ Kavli Institute for the Physics and Mathematics of the Universe, 5-1-5 Kashiwanoha, Kashiwa 277-8583, Japan \and
\small$^{9}$ Division of Science, National Astronomical Observatory of Japan, 2-21-1 Osawa, Mitaka, Tokyo 181-8588, Japan \and
\small$^{10}$ Astrophysics, University of Oxford, Denys Wilkinson Building, Keble Road, Oxford OX1 3RH, United Kingdom \and
\small$^{11}$ University of California Observatories, 1156 High Street, Santa Cruz, CA 95064, USA \and
\small$^{12}$ Center for Interdisciplinary Exploration and Research in Astrophysics (CIERA) and Department of Physics \and 
\small{and Astronomy, Northwestern University, Evanston, IL 60208, USA} \and
\small$^{13}$ International Centre for Radio Astronomy Research, Curtin University, Turner Avenue, Bentley, WA 6102, Australia \and
\small$^{14}$ LPC2E, OSUC, Univ Orleans, CNRS, CNES, Observatoire de Paris, Orleans F-45071, France \and
\small$^{15}$ Department of Physics, College of Natural and Mathematical Sciences, University of Dodoma, \and
\small{1 Benjamin Mkapa Road, 41218 Iyumbu, Dodoma 259, Tanzania}\and
\small$^{16}$ Department of Astronomy and Astrophysics, University of Chicago, Eckhardt, \and
\small{5640 S Ellis Ave, Chicago, IL 60637, USA} \and
\small$^{17}$ Max-Planck-Institut für Radioastronomie, Bonn 53121, Germany \and
\small$^{18}$ Department of Mathematical Sciences, University of South Africa, Cnr Christiaan de Wet Rd \and
\small{and Pioneer Avenue, Florida Park, Roodepoort 1709, South Africa}\and
\small$^{19}$ Cahill Center for Astronomy and Astrophysics, MC 249-17 California Institute of Technology,\and
\small{Pasadena, 91125, USA}\and
\small$^{20}$ School of Mathematical and Physical Sciences, Macquarie University, Sydney, NSW 2109, Australia \and
\small$^{21}$ Australia Telescope National Facility, CSIRO, Space \& Astronomy, PO Box 76, Epping, NSW 1710, Australia \and
\small$^{22}$ Astrophysics and Space Technologies Research Centre, Macquarie University, Sydney, NSW 2109, Australia \and
\small$^{23}$ ASTRO3D: ARC Centre of Excellence for All-sky Astrophysics in 3D, Canberra, ACT 2601, Australia \and
\small$^{24}$ Center for Astrophysics, Harvard \& Smithsonian, 60 Garden Street, Cambridge, MA 02138, USA \and
\small$^\ast$Corresponding author. Email: manisha.caleb@sydney.edu.au
}

\begin{document} 
\maketitle

\begin{abstract} \bfseries \boldmath
Fast radio bursts (FRBs) are enigmatic millisecond-duration signals which encode otherwise unattainable information on the plasma which permeates our universe, providing insights into magnetic fields and gas distributions. Here we report the discovery of FRB~20240304B originating at redshift $2.148\pm0.001$ corresponding to just 3 billion years after the Big Bang. FRB~2024030 was detected with the MeerKAT radio telescope and localized to a low-mass, clumpy, star forming galaxy using the James Webb Space Telescope. This discovery doubles the redshift reach of localized FRBs and probes ionized baryons across $\approx80\%$ of cosmic history. Its sightline, intersecting the Virgo Cluster and a foreground group, reveals magnetic field complexity over many gigaparsec scales. Our observations establish FRB activity during the peak of cosmic star formation and demonstrate that FRBs can probe galaxy formation during the most active era in cosmic time.
\end{abstract}

\noindent
The enigmatic FRB phenomenon is characterised by bright, coherent, millisecond-duration radio pulses of as-yet-unidentified origin(s) \cite{lorimer+2007, thornton+2013}. The pulses exhibit a frequency-dependent delay in the arrival time at the telescope consistent with propagation through cold ionized plasma. This time delay provides a direct estimate of the integrated electron column densities along the lines-of-sight to these bursts, called dispersion measure (DM). When coupled with spectroscopic redshifts determined from their associated host galaxies, it becomes possible to trace the distribution of ionized matter across the large-scale structure of the Universe through the `Macquart relation' \cite{macquart+2020}. 
This relation has enabled the first direct measurement of the missing baryons and, consequently, provided astronomers with a novel means of `weighing' the Universe \cite{macquart+2020}. 
The Macquart relation also offers independent and complementary constraints on the Hubble constant H$_0$ \cite{jgp+22}, baryonic feedback processes in galaxies \cite{bpm+24}, and even has the potential to test scenarios of helium and hydrogen reionization \cite{cfs19, bkm+21}, all of which are highly valuable tests of the standard cosmological model and galaxy evolution in our early Universe. However, it is not known whether FRBs can be produced at the high redshifts required to test some of these models and scenarios. Certain progenitor scenarios link FRBs to young neutron stars born in supernovae or gamma‑ray bursts \cite{zhang2014, cw2016, mbm2017}, implying that FRB birth rates could trace the cosmic star‑formation history.
In this context, detecting FRBs at high redshifts would offer a powerful probe of galaxy evolution and stellar populations during the peak epochs of star formation as well as their impact on the intergalactic medium. 

Of the $\sim100$ FRBs with redshifts determined from their host galaxies, the vast majority reside at relatively low redshifts of $z\lesssim0.5$ (e.g., \cite{gfk+23, sharma+24}) and the distribution at  $z\gtrsim 1$ \cite{md2014} remains uncharted territory. This gap is driven by the sensitivity horizon of current FRB detection facilities and the challenges of obtaining host galaxy spectra at these high redshifts using ground-based telescopes. 
Our discovery of \frb{}\, at \redshift{}\, during the peak of galaxy formation approximately ten billion years ago underscores the potential of FRBs as powerful probes of the cosmic web.


\subsection*{Discovery and observations of FRB~20240304B}

\begin{figure}
\centering
  \includegraphics[width=0.6\textwidth]{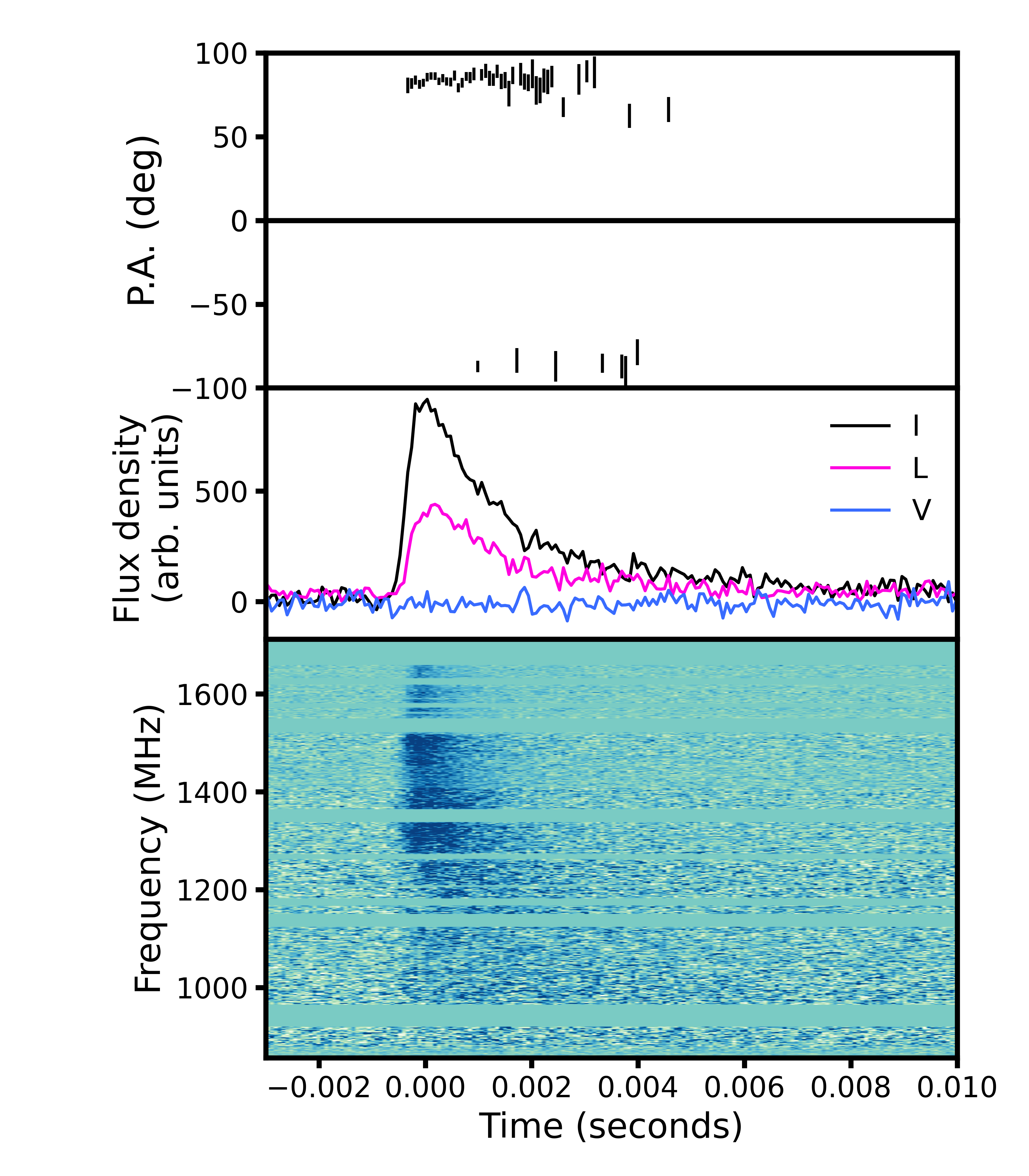}
    \caption{\textbf{Dynamic spectrum and polarization profile of FRB~20240304B}. The data have a time resolution of 38.28~$\mu$s and are coherently de-dispersed to a DM of 2458.20\,$\dmunits$ (bottom panel) and corrected for an RM of $-55.6 \, \rmunits$. The horizontal gaps in the bottom panel, where data are missing, are a result of radio frequency interference excision during data processing. The top two panels show the polarisation position angles. The middle panel shows the Stokes parameter pulse profiles at 1284 MHz where black represents the total intensity, magenta represents linear polarization and blue represents circular polarization. The data are uncalibrated and the flux density is in arbitrary units.}
\label{fig:frb-dynspec}
\end{figure}

\frb{} was discovered by the MeerTRAP backend instrument (TUSE) (e.g., \cite{rbc+22}) at the MeerKAT radio telescope on 04 March 2024 during an open-time observation (Proposal ID: SCI-20230907-FD-01). It was detected at 00:50:12.567 UTC with an observed \dmfrb\ of $2462.48\,\dmunits$, which we refine to $2458.20\pm0.01\,\dmunits$ after accounting for the temporal pulse broadening due to the effect of scattering \cite{methods}. Offline processing \cite{methods} localised the burst to RA=12:11:59.29$\pm0.28\arcsec$ and Dec=$+$11:48:46.86$\pm0.48\arcsec$ (see Figure \ref{fig:on_off}) with a S/N of 114.7, and revealed the time-frequency structure of the burst (Figure \ref{fig:frb-dynspec}).

The total \dmfrb\ of an FRB is the sum of various line-of-sight contributions from the Milky Way, the cosmos and the host galaxy, i.e. $\dmfrb = \dmmwism + \dmhalo + \dmcosmic + \dmhost/(1+z)$, where \dmcosmic\ is the sum of contributions from gas in the circumgalactic medium (CGM) of potential intervening galaxy halos, and the diffuse intergalactic medium (IGM) in between. We estimate $ \dmmwism = 28\,\dmunits\,$ \cite{ne2001} and $ \dmhalo = 40\,\dmunits$ \cite{PZ19,cook+23} with a systematic error of a few tens \dmunits. This implies a cosmic \dmcosmic\ of 2330 \,\dmunits, suggesting a high-redshift  origin for the FRB, with a predicted mean redshift of $z_\mathrm{Macquart} = 2.8^{+0.6}_{-1.2}$ based on the Macquart relation \cite{macquart+2020}.


\subsection*{Host galaxy identification and redshift}

Consistent with a high-redshift origin, archival searches of public surveys with relatively shallow flux limits did not reveal a host galaxy counterpart at the FRB's position. We therefore carried out ground-based optical and infrared follow-up observations of \frb{} with Keck/LRIS and MMT/MMIRS, respectively \cite{methods}. Despite very deep optical
imaging at the radio coordinates of \frb{}, 
no host galaxy was detected to $5\sigma$ limiting magnitudes of $R = 26.24$ AB and $J=23.38$ AB. 

\begin{figure}
\centering
  \includegraphics[width=0.76\textwidth]{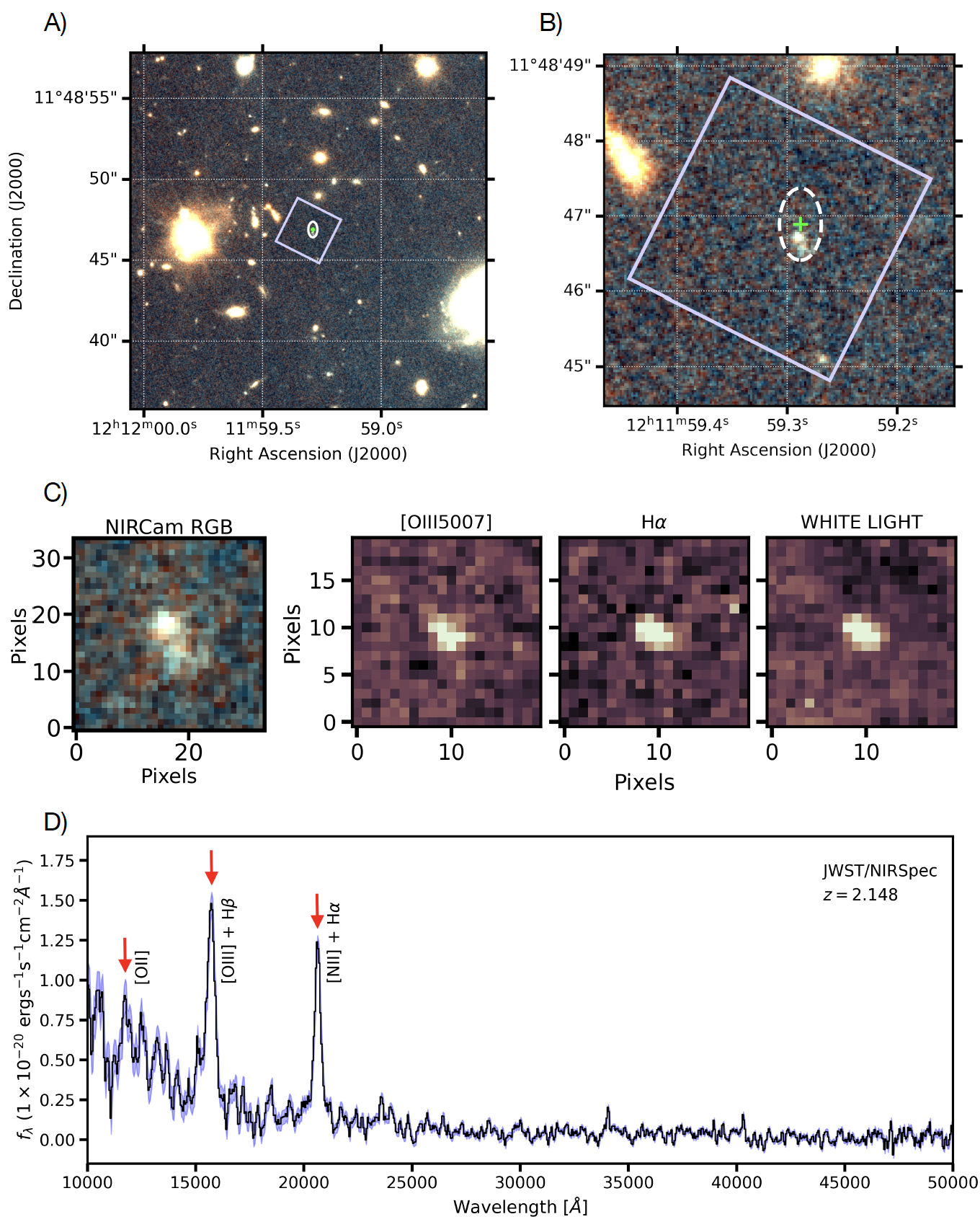}
    \caption{\textbf{RGB composite NIRCam image and spectrum of the FRB host galaxy at \redshift{}}. A) The image in the top left provides a larger-scale view of the field. B) The image in the top right shows a $4.6\arcsec \times 4.6\arcsec$ NIRCam image cutout with a $0.28\arcsec \times 0.48 \arcsec$ localization uncertainty (dashed white circle) for the FRB position. The green cross marks the FRB's best-fit position. The IFS field of view is outlined by the purple square, encompassing the host. C) The various panels display cutouts of the NIRCam RGB image of the host galaxy. [OIII5007], H$\alpha$, and white-light images extracted from the IFS data, show the presence of star-forming emission at the FRB location. D) The bottom panel showcases the NIRSpec spectrum of the host galaxy with various emission lines highlighted by the arrows. The shaded region represents the error on the spectrum.}
\label{fig:jwst_loc}
\end{figure}

Given the absence of a detectable host in our ground-based observations, we obtained a director's discretionary time observation (PID 6779) with the James Webb Space Telescope (JWST) using its Near Infrared Camera (NIRCam) and Near Infrared Spectrograph (NIRSpec) Integral Field Spectrograph (IFS), to conduct a deeper search for a faint or distant host galaxy. NIRCam imaging of the field with the F200W filter in the short wavelength band and the F322W2 filter in the long wavelength band revealed a putative host galaxy separated by $\sim0.3\arcsec$ from the FRB position (Figure \ref{fig:jwst_loc}). We measure the galaxy's magnitudes to be $m_{F200W} = 27.82\pm0.06$ (AB) and $m_{F322W2} = 28.72\pm0.03$ (AB) in the F200W and F322W2 filters, respectively. Adopting the Probabilistic Association of Transients to their Hosts (PATH;\cite{path}) formalism to assess the posterior probability $\pox$ of  associating the FRB to this candidate, we recover a robust association of $\pox=\pathpost$.
The remainder of the probability is attributed to an unseen host galaxy with $\pux = \puxpost$ \cite{methods}. NIRSpec observations of the host allowed us to identify strong \Halpha\, and \OIII\, emission lines in the spectra from which we measure the redshift to be \redshift{} (Figure \ref{fig:jwst_loc}). This confirms that \frb{} is the most distant FRB detected to date and the first to be detected at cosmic noon.

At this redshift, the rest-frame optical luminosity of the host is $L \approx 1.9 \times 10^{9}\,L_{\odot}$. This is $\sim 10$ times more luminous than the host of FRB\,20121102A \cite{tbc+17}, but still among the lowest known optical luminosities for the host of any FRB to date.


\subsection*{Host galaxy properties}

We estimate the stellar mass of the host galaxy to be $\sim 10^{7}$ M$_\odot$ with a star formation rate (SFR) of $\sim0.2$\,$\sfunits$ and gas-phase metallicity of $\sim10-20\%$ solar metallicity \cite{methods}. 
The specific star formation rate (sSFR; star formation normalised by stellar mass) places the galaxy above the star forming main sequence (see Figure \ref{fig:sfr_mass_metallicity}) indicating that the galaxy has a higher SFR than what 
is typical for its low stellar mass \cite{speagle+2014, topping+21, popesso+23, clarke+24}. The host galaxy of \frb{} could therefore be a low-metallicity, starburst, dwarf galaxy in an early evolutionary phase. Assuming a constant star formation rate over its lifetime, the galaxy’s stellar formation timescale, $t_{\rm form} \equiv M_{*}/{\rm SFR} = 51.7 \, \rm Myr$, suggesting that it is very young and could have formed its entire stellar mass in a relatively short time span \cite{methods}.  
Overall, these stellar mass and metallicity values are atypical of the known FRB host galaxy population \cite{gfk+23, sharma+24}. However, we note that very few FRB host galaxies have been observed at such faint magnitudes at any redshift.

\subsection*{Clues to FRB~20240304's origin from its host galaxy}

Our measured redshift is consistent with the DM-estimated range of [1.628, 3.397] (95\% confidence interval) according to the Macquart relation \cite{macquart+2020, Hoffmann2025}, and corresponds to a luminosity distance D$_{L}$ of 17.4~Gpc. The redshift measurement allows us to establish the rest-frame energy scale of \frb{} to be $\sim10^{41}$~erg with an isotropic-equivalent energy density of $\sim10^{32}$~erg Hz$^{-1}$ for a S/N of 114.7 (see \cite{methods} and Figure \ref{fig:shannonplot}). Since the intrinsic spectrum of the FRB is unknown, we do not apply a K-correction. While this introduces some uncertainty, our main conclusions remain unaffected for typical FRB spectral indices \cite{macquart+2019}. 
Our measured energy is on the higher end of the typical FRB isotropic-equivalent burst energies ($10^{36-41}$ erg; \cite{kirsten+2024}) and is comparable to the non-K-corrected burst energy of FRB~20220610A of $(6.4\pm0.7) \times 10^{32} $ erg Hz$^{-1}$ at $z \sim 1$ \cite{rbb+22}.

Beyond energetics, the host galaxy provides crucial insights into the nature of the progenitor. The low stellar mass and ongoing star formation in the host galaxy of \frb{} favour a short delay time between star formation and FRB production. These properties are more naturally explained by progenitor models involving prompt channels -- those in which the FRB source forms within a few million years of when star formation occurred -- such as young magnetars \cite{brd21}. In contrast, progenitors requiring long delay times, such as neutron star mergers, would typically be associated with older stellar populations and evolved galaxies (e.g., \cite{grw+2020, nugent+22}). Furthermore, in low-metallicity environments, the initial mass function favours the formation of more massive stars, and thus magnetars \cite{hm2025, hl2009, Kroupa01, mkd+2012}. This is consistent with recent evidence linking FRBs to star-forming galaxies \cite{gfk+23} and the detection of FRB-like bursts from the Galactic magnetar SGR~J1935$+$2154 \cite{bochenek+20, chime_sgr1935}. 
Finally, our detection of the first FRB near the epoch of peak cosmic star formation strengthens the case that most FRBs likely have short delay times relative to star formation. If most FRBs had significant $\gtrsim$~Gyr delay times, they would be much more difficult to detect at high redshift due to the lower intrinsic rates per unit volume. 


\subsection*{Tracing the magnetic fields in the FRB's line of sight}

\frb{} is highly linearly polarized with a linear polarization fraction of 49\% and a negligible circular polarization fraction of 3\%. The wavelength-dependent variation in the linear polarization position angle due to ordered magnetic fields in astrophysical plasmas is referred to as Faraday rotation. The magnitude of this effect is quantified by the rotation measure (RM), which reflects the line-of-sight integral of the electron density weighted by the strength and direction of the magnetic field. The pulse exhibits Faraday rotation, with an RM of $-55.6\pm0.5\,\rmunits$ \cite{methods}. In contrast, the closest known pulsar along the line-of-sight, B1133$+$16, is 9.6~deg away and, has an RM of $2.56\pm3$\, \rmunits\, \cite{kjk+24}, while the the closest extragalactic source from the NVSS VLA Sky Survey catalog, is 0.2~deg away with an RM of $-6.4\pm7.9$\,\rmunits\, \cite{tss2009}. The contribution from the smoothed Galactic foreground is $0.12\pm7.42$\,\rmunits \cite{hab+22}. 
Overall, this indicates a low foreground contribution implying that the observed RM is intrinsic to the FRB's host galaxy, local environment, and/or the intervening IGM. 
If the RM is intrinsic to the local environment, its rest-frame value would be amplified by a factor of $(1+z)^2 \approx 10$, yielding an intrinsic RM of approximately $\sim -550$\,\rmunits. This is relatively high in the context of the known FRB population, where many RMs are consistent with low or modest magnetization \cite{mpp+23, sherman+23, mckinven+23} (see Figure \ref{fig:DMvsRM}). Given the large DM and the comparatively low foreground RM, the average line-of-sight magnetic field is extremely small ($–28$~nG). 
The relatively low observed absolute RM of the FRB implies that the magnetic fields along the line-of-sight are either weaker than expected or most likely more complex, with cancellations \cite{kbm+08} or reversals of the field direction.


\subsection*{Foreground contributions to the DM}

We use the properties of \frb{} to extend the Macquart relation to $z >2$ (Figure \ref{fig:zdm}) and find that the FRB lies above the median \dmcosmicfrb, indicating an excess ionized electron column density of approximately $\approx550$\,\dmunits\, along the line-of-sight. This excess could arise from the IGM, halos of intervening galaxies, and/or contributions from the host galaxy, including the FRB's local circumburst environment. We find that the FRB sightline intersects the Virgo Cluster, passing about $\sim 1$\,Mpc from its centre and $\sim240$~kpc from a galaxy group we identified at $z=0.31131$ \cite{methods}. We estimate the Virgo Cluster contributes $\sim235$\,\dmunits\ and the foreground galaxy group $\sim33$\,\dmunits\ to the FRB’s dispersion measure \cite{methods}, suggesting that these structures account for the majority of the observed DM excess. This result could also inform tomographic reconstructions of the cosmic web along this sight-line \cite{simha+2020,khrykin+2024}, which we defer for future work.


\subsection*{``Smashing the $z\gtrsim2$ Window"}

Given the substantial leap in redshift  with our discovery, one might ask: why did the association of the first FRB to a $z>2$ galaxy occur $\sim 8$~years and $\sim 100$ well-localized FRBs after the first high confidence association \cite{clw+17}?
Related, is this an extremely rare (i.e, lucky) $O(1\%)$ event? Intriguingly, the answer is that the result is  perfectly consistent with what we have learned from the $\sim 100$~FRB associations and their host galaxies.   

\begin{figure}
\centering
  \includegraphics[width=0.9\textwidth]{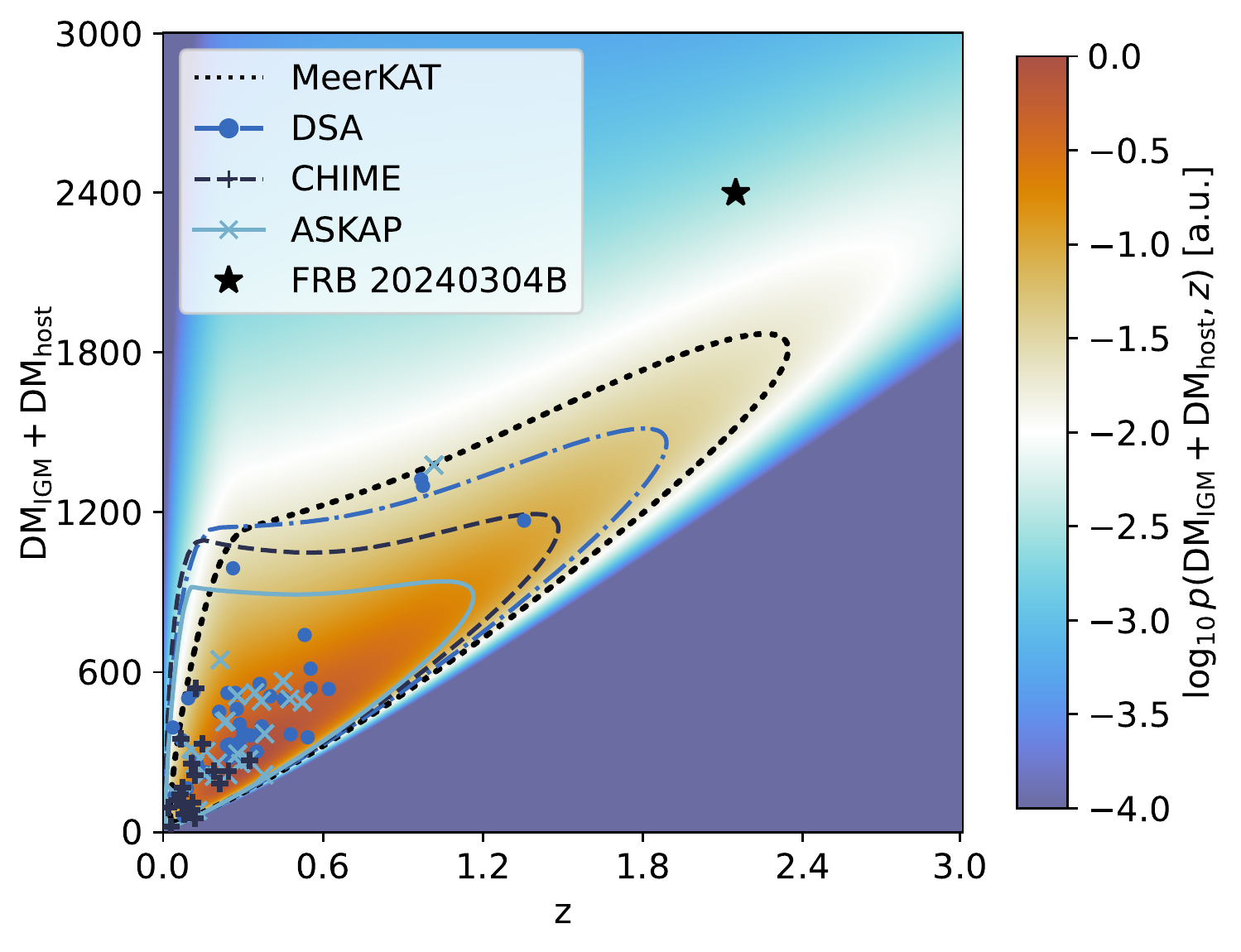}
    \caption{\textbf{Predicted $z$-DM distribution for MeerKAT using best-fit FRB parameters from \cite{Hoffmann2025} and the \textsc{zDM} code \cite{jpm+22a} (uncertainties in those parameters are not accounted-for)}. The overplotted contours encompass 90\%\ of the integrated probability for the ASKAP/CRAFT (solid light blue), DSA (dash-dot blue), CHIME-FRB (dashed dark blue), and MeerKAT/MeerTRAP (dotted black) experiments. Light blue crosses show localised FRBs from ASKAP/CRAFT \cite{shannon+24}, blue circles show localised FRBs from DSA \cite{sharma+24} and dark blue crosses show localised FRBs from CHIME/FRB \cite{chime25}. The black star is \frb{}.
    Only MeerKAT has the sensitivity such that $>5\%$ of its
    detected FRBs are predicted to occur at $z>2$.
    }
\label{fig:zdm}
\end{figure}

First and foremost, the projects which have dominated the discovery of these FRBs have insufficient sensitivity to detect $z>2$~FRBs. This statement hinges on the energetics of the FRB population which has been estimated from the well-localized population \cite{jgp+22, src+24}. While the FAST telescope has the sensitivity to detect FRBs up to $z \sim 10$ under ideal conditions \cite{zhang2018}, it lacks the high-precision localization required to associate FRBs to host galaxies. 
Figure~\ref{fig:zdm} shows the contours encompassing 90\%\ of the integrated probability of FRB detection in the $z$-DM plane for the Australian Square Kilometer Array Pathfinder (ASKAP/CRAFT; \cite{bdp+19}), Deep Synoptic Array (DSA; \cite{ravi2019}), Canadian Hydrogen Intensity Mapping Experiment (CHIME-FRB; \cite{mbb+23}), and MeerKAT/MeerTRAP \cite{rbc+22} experiments with the full set of published FRBs detected to date for comparison, illustrating that MeerKAT is the only current detector with the sensitivity to the $z>2$ universe.

Even for the DSA, which is the next most sensitive of these experiments, we expect $<5\%$ of the FRBs to occur at $z>1.6$, and $<3\%$ at $z>2$. We emphasize however, as others have before us (see also \cite{zhang2018, Hoffmann2025}), that these surveys are still expected to detect FRBs with $\dmfrb > 2000\,\dmunits$ -- and most have --  but these large values are due to the combination of gas in/around the host galaxy and/or foreground structures (e.g. galaxy clusters; \cite{lks+23}). In contrast, we predict that 6.3\% of the FRBs from the MeerTRAP coherent beam survey will occur at $z>2$.

The identification of an FRB at \redshift{} marks a critical milestone in FRB astronomy. By identifying the first FRB at cosmic noon, we demonstrate that at least some FRBs are not produced by formation channels involving significant delay times. Taken together, the low stellar mass, active star formation, and low metallicity of the host galaxy 
favour a magnetar origin.
The forthcoming era of high-sensitivity radio interferometeric FRB science with facilities such as the SKAO’s Mid telescope in conjunction with space-based imaging and spectroscopy will build on these results, providing a high-redshift census of FRBs, dramatically improving our understanding of their progenitors, and cementing FRBs as essential cosmological probes.



\begin{table}
\centering
\caption{\textbf{Observed and inferred properties of FRB~20240304B.}}
\label{tab:burstproperties}
\begin{tabular}{cc}

\hline
\multicolumn{2}{c}{Measured burst properties}\\
\hline
MJD$_\text{topo}^\text{a}$  & 60373.034867670896 \\
UTC$_\text{topo}^\text{a}$  & 2024-03-04 00:50:12.56\\
Beam                        & Coherent beam \\
RA                          & 12:11:59.29$\pm0.28\arcsec$ \\
Dec                         & $+$11:48:46.86$\pm0.48\arcsec$  \\
$l$                         & 269.8676119~deg  \\
$b$                         & 72.1035378~deg\\
Scattering-corrected DM     & $2458.20\pm0.01\,\dmunits$\\
$\text{Discovery S/N}$ (856--1712 MHz)  & 39 \\
$\text{TB data S/N}$ (856--1712 MHz)    & 114.7 \\
Scattering time, $\tau_\mathrm{s}$ at 1~GHz & $5.6\pm0.3$ \text{ms}\\
Rotation measure, RM        & $-55.6\pm0.5$ \text{rad} \text{m$^{-2}$} \\
Linear polarization fraction, $L/I$     & $0.49\pm0.01$\\
Circular polarization fraction, $|V|/I$ & $0.03\pm0.01$ \\

\hline  
\multicolumn{2}{c}{Inferred burst properties}\\
\hline 
$S_\text{peak}^\text{b}$     & $0.49\pm0.01$ Jy\\
$F_{\nu}$                    & $2.75\pm0.05$ Jy ms\\
DM$_{\text{MW,NE2001}}$      & 28.1 $\text{pc} \: \text{cm}^{-3}$\\
DM$_{\text{MW,YMW16}}$       & 20.8 $\text{pc} \: \text{cm}^{-3}$\\
DM$_{\text{halo}}$           & 40 $\text{pc} \: \text{cm}^{-3}$\\
DM$_{\text{Virgo Cluster}}$  & 235 $\text{pc} \: \text{cm}^{-3}$\\
DM$_{\text{galaxy group, z=0.31131}}$  & 33 $\text{pc} \: \text{cm}^{-3}$\\\\
\hline 
\multicolumn{2}{c}{$^\text{a}$ Topocentric arrival times measured at the highest frequency channel, 1711.58~MHz and.}\\
\multicolumn{2}{c}{$^\text{b}$ Estimated for the TB data S/N.}\\

\end{tabular}
\end{table}

\begin{table}
\centering
\caption{\textbf{Observed and inferred properties of the host galaxy.}}
\label{tab:burstproperties}
\begin{tabular}{cc}

\hline
\multicolumn{2}{c}{Host galaxy properties}\\
\hline
RA$_\mathrm{host}$   & 12:11:59.29 \\
DEC$_\mathrm{host}$   & $+$11:48:46.63 \\
Redshift ($z_\mathrm{spec}$)  & $2.148\pm0.001$ \\
F200W (AB mag) & $27.82\pm0.06$\\
F322W2 (AB mag) & $28.73\pm0.03$ \\
Rest-frame optical luminosity, $L$ &  $1.9 \times 10^{9}\,L_{\odot}$ \\
Stellar metallicity (\fastpp, Z$_*$/Z$_{\odot}$) & $0.2 \pm 0.0$ \\
Stellar mass,  log (\fastpp, M$_{*}$/M$_{\odot}$) & $6.89^{+0.06}_{-0.01}$ \\
\fastpp, SFR$_{100\mathrm{Myr}}$ & 0.15 M$_{\odot}$ yr$^{-1}$\\
f(H$_\alpha$ + \NII) & $(4.2 \pm 0.2)\times 10^{-18}$ erg s$^{-1}$ cm$^{-2}$\\

\hline

\end{tabular}
\end{table}


\clearpage 

%
\bibliography{science_template} 
\bibliographystyle{sciencemag}

%
%
%
%
%
%


\section*{Acknowledgments}
M.C. would like to thank Elaine Sadler and Adam Deller for discussions, and Ryan Shannon for providing the data to reproduce Figure \ref{fig:shannonplot}. 
This work is based on observations made with the NASA/ESA/CSA James Webb Space Telescope. The data were obtained from the Mikulski Archive for Space Telescopes at the Space Telescope Science Institute, which is operated by the Association of Universities for Research in Astronomy, Inc., under NASA contract NAS 5-03127 for JWST. These observations are associated with the DD program 6779 (PI Caleb). The MeerKAT telescope is operated by the South African Radio Astronomy Observatory (SARAO), which is a facility of the National Research Foundation, itself an agency of the Department of Science and Innovation. All the authors thank the MeerKAT LSP teams for allowing commensal observing and the staff at SARAO for scheduling MeerKAT observations. MeerTRAP (PI Stappers) observations use the FBFUSE and TUSE computing clusters for data acquisition and storage. The MeerTRAP collaboration acknowledges funding from the European Research Council under the European Union’s Horizon 2020 research and innovation programme (grant agreement No 694745). The FBFUSE equipment was designed, funded and installed by the Max-Planck Institut f{\"u}r Radioastronomie (MPIfR) and the Max-Planck-Gesellschaft. Parts of the analysis were performed using the Nan\c{c}ay Data Centre computing facility (CDN -- Centre de Donn\'{e}es de Nan\c{c}ay). The CDN is hosted by the Observatoire Radioastronomique de Nan\c{c}ay in partnership with Observatoire de Paris, Universit\'{e} d'Orl\'{e}ans, OSUC and the CNRS. The CDN is supported by the R\'{e}gion Centre Val de Loire, d\'{e}partement du Cher. MMT Observatory access was supported by Northwestern University and the Center for Interdisciplinary Exploration and Research in Astrophysics (CIERA). Observations reported here were obtained at the MMT Observatory, a joint facility of the University of Arizona and the Smithsonian Institution. Some of the data presented herein were obtained at the W. M. Keck Observatory, which is operated as a scientific partnership among the California Institute of Technology, the University of California and the National Aeronautics and Space Administration. The Observatory was made possible by the generous financial support of the W. M. Keck Foundation. The authors wish to recognize and acknowledge the very significant cultural role and reverence that the summit of Maunakea has always had within the indigenous Hawaiian community. We are most fortunate to have the opportunity to conduct observations from this mountain.

\paragraph*{Funding:}
This project has received funding from the European Research Council (ERC) under the European Union’s Horizon 2020 research and innovation programme (grant agreement no. 694745).
M.C. acknowledges support of an Australian Research Council Discovery Early Career Research Award (project number DE220100819) funded by the Australian Government.
J.T. and B.W.S. acknowledge funding from an STFC Consolidated grant. 
I.P.M. acknowledges funding from an NWO Rubicon Fellowship, project number 019.221EN.019. 
T.N. and K.G. acknowledge support from Australian Research Council Laureate Fellowship FL180100060. 
Parts of this research were conducted by the Australian Research Council Centre of Excellence for Gravitational Wave Discovery (OzGrav), project number CE230100016. 
J.X.P., A.C.G., L.M.R. acknowledge support from NSF grants AST-1911140, AST-1910471 and AST-2206490 as members of the Fast and Fortunate for FRB Follow-up team. 
A.C.G. and the Fong Group at Northwestern acknowledges support by the National Science Foundation under grant Nos. AST-1909358, AST-2206494, AST-2308182 and CAREER grant No. AST-2047919.
N.T. acknowledges support from FONDECYT grant 1252229.
I.S.K. and N.T. acknowledge support from grant ANID / FONDO ALMA 2024 / 31240053
S.S. is a joint NU-UC Brinson Postdoctoral Fellow supported by the Brinson Foundation.
W.F. gratefully acknowledges support by the David and Lucile Packard Foundation, the Alfred P. Sloan Foundation, the Research Corporation for Science Advancement through Cottrell Scholar Award \#28284, and the National Science Foundation.
C.D.M knowledges support from the NASA Grant 80NSSC22K1649.

\paragraph*{Author contributions:}
M.C. drafted the manuscript with input from co-authors and is the PI of the JWST data.
B.W.S. is the PI of MeerTRAP.
M.C., B.W.S., K.R., and F.J. reduced the MeerKAT TUSE/TB data for scattering and polarization analyses.
M.C. and B.W.S. interpreted the burst energetics and their implications for the FRB emission mechanism.
I.P.M. and L.N.D. performed the localisation and astrometry of the MeerKAT radio data.
T.N. and K.G. designed the JWST observing program.
T.N. conducted the JWST data reduction, analysis, and spectral study of the host galaxy.
S.S., L.M.R. and J.X.P. obtained the Keck/LRIS observations.
C.D.M., Z.L. and X.D. performed the KCWI observations.
I.S.K. and N.T. carried out the Keck/LRIS and KCWI data reduction.
I.S.K., N.T., M.C., and B.W.S. analysed the foreground DM contribution.
A.G. and W.F. designed the observations and processed the MMT/MMIRS J-band images.
L.K. and J.X.P. performed the host association.
T.N., K.G., J.X.P., and N.T. interpreted the host-galaxy properties.
C.W.J. and J.H. conducted the zdm analyses.
T.B. performed the initial FRB localisation using MeerKAT TUSE data.
E.B. built and designed the beamformer used by MeerTRAP and the complex channelised data capture system. 
M.K. provided user equipment and resources to enable MeerKAT observations.
All co-authors contributed to the manuscript through comments and text.

\paragraph*{Competing interests:}
There are no competing interests to declare.

\paragraph*{Data and materials availability:}
JWST data used in this analysis is publicly available through the Mikulski Archive for Space Telescopes. All code necessary for analyses of the data are available on GitHub and Zenodo:


\subsection*{Supplementary materials}
Materials and Methods\\
Supplementary Text\\
Figs. S1 to S7\\
Table S1\\
References \textit{(57-\arabic{enumiv})} 



\newpage

\renewcommand{\thefigure}{S\arabic{figure}}
\renewcommand{\thetable}{S\arabic{table}}
\renewcommand{\theequation}{S\arabic{equation}}
\renewcommand{\thepage}{S\arabic{page}}
\setcounter{figure}{0}
\setcounter{table}{0}
\setcounter{equation}{0}
\setcounter{page}{1} 


\begin{center}
\section*{Supplementary Materials for\\ \scititle}

Manisha Caleb$^{1,2\ast}$, 
Themiya	Nanayakkara$^3$,
Benjamin Stappers$^4$, \\
In\'es Pastor-Marazuela$^4$,
Ilya S. Khrykin$^5$,
Karl Glazebrook$^{3}$, 
Nicolas	Tejos$^5$, \\
J. Xavier Prochaska$^{6,7,8}$,
Kaustubh Rajwade$^9$,
Lluis Mas-Ribas$^{6,10}$, \\
Laura N. Driessen$^1$, 
Wen-fai	Fong$^{11}$,
Alexa C. Gordon$^{11}$, 
Jordan Hoffmann$^{12}$, \\
Clancy W. James$^{12}$,
Fabian Jankowski$^{13}$, 
Lordrick Kahinga$^{6,14}$, \\
Michael	Kramer$^{15,4}$, 
Sunil Simha$^{11,15}$,
Ewan D. Barr$^{16}$,  \\
Mechiel Christiaan Bezuidenhout$^{17}$,
Xihan Deng$^{18}$,
Zeren Lin$^{18}$,  \\
Lachlan Marnoch$^{19, 20, 21, 22}$, 
Christopher D. Martin$^{18}$,
Anya Nugent$^{23}$, \\
Kavya Shaji$^{1,20}$,
Jun Tian$^4$

\small$^\ast$Corresponding author. Email: manisha.caleb@sydney.edu.au
\end{center}

\subsubsection*{This PDF file includes:}
Materials and Methods\\
Supplementary Text\\
Figures S1 to S7\\
Table S1 


\newpage


\clearpage

\renewcommand\thesection{S\arabic{section}} 
\setcounter{section}{0} 

\renewcommand\thetable{S\arabic{table}} 
\setcounter{table}{0} 

\renewcommand\thesection{S\arabic{section}} 
\setcounter{section}{0}

\renewcommand\theequation{S\arabic{equation}}
\setcounter{equation}{0}

\pagenumbering{arabic}
    \setcounter{page}{1}

\section{Materials and Methods}

\subsection{MeerKAT Radio Observations}\label{sec:meerkat}

The MeerKAT radio telescope is a 64-dish interferometer operated by the South African Radio Astronomy Observatory (SARAO) in the Karoo region in South Africa. The dishes are spread over 8 km, with 40 of them concentrated in the inner $\sim1$-km core. The MeerTRAP project is a commensal programme to search for pulsars and fast transients whilst piggy-backing on other science programmes at MeerKAT.  In the observations presented in this work, MeerKAT operated at a centre frequency of 1284~MHz (L-band) with a usable bandwidth of $\sim770$~MHz. The MeerTRAP backend is the association of two systems: the Filterbank and Beamforming User Supplied Equipment (FBFUSE), a multi-beam beamformer that was designed and developed at the Max-Planck-Institut f{\"u}r Radioastronomie in Bonn \cite{barr2018, CBK+21}, and the Transient User Supplied Equipment (TUSE), a real-time transient detection instrument developed by the MeerTRAP team.
When operating at L-band in the coherent mode, the data from the inner 40 dishes of the $\sim1$-km core of the array are coherently combined to form up to 780 coherent beams (CBs) on the sky with an aggregate field-of-view (FoV) of $\sim0.4$~deg$^{2}$ (i.e. overlap at 25\% of the peak power). In the incoherent mode, the intensities of all available MeerKAT dishes (up to a maximum of 64) are added to create a less sensitive but much wider FoV of $\sim$1~deg$^{2}$ \cite{rsw+21} beam. 
\frb\ was detected during an open-time observation (Proposal ID: SCI-20230907-FD-01) \cite{de_gasperin_victoria_2025}.
Its detection triggered the storage of channelized voltages around the time of the burst from the 62 antennas included in the observations. The transient buffer (TB) saved 300~ms of data around the time of the burst by taking into account the time delay due to dispersion. These data were used to produce visibilities which in turn were used to create radio images and localise the burst (see section below) \cite{rdb+24}. We then used the refined positional coordinates and channelized voltages to form a coherent beam at this position by multiplying the voltages with the appropriate weights as a function of antenna and frequency. This resulted in a dataset with all Stokes parameters and an effective time resolution of 4.789~$\mu$s across 4096 frequency channels.

\subsection{FRB~20240304B localization and astrometry} \label{sec:localisation}

To localize the burst in the TB data, we followed the procedure outlined in \cite{rdb+24}. In summary, the full observing bandwidth is divided into 64 frequency subbands, each assigned a different start time to account for the dispersion delay of the burst. 
Since the TB data were only saved for 56 subbands we only use those for the localization.
For each subband, we first correlated the data using \texttt{xGPU} \cite{clark_accelerating_2011} and converted it into a measurement set (MS) using \texttt{DiFX} \cite{deller_difx_2007, deller_difx-2_2011} and \texttt{CASA} \cite{mcmullin_casa_2007}. We then produced images for each of the MSs using \texttt{WSClean} \cite{offringa_wsclean_2014}. After visually inspecting these images, we flagged those affected by significant RFI and averaged the remaining unflagged images to create a single, full-bandwidth image. After removing subbands that were either missing or affected by RFI, we used a total of 31 subbands out of the 56 available for the final image.

\begin{figure}
\centering
  \includegraphics[width=0.9\textwidth]{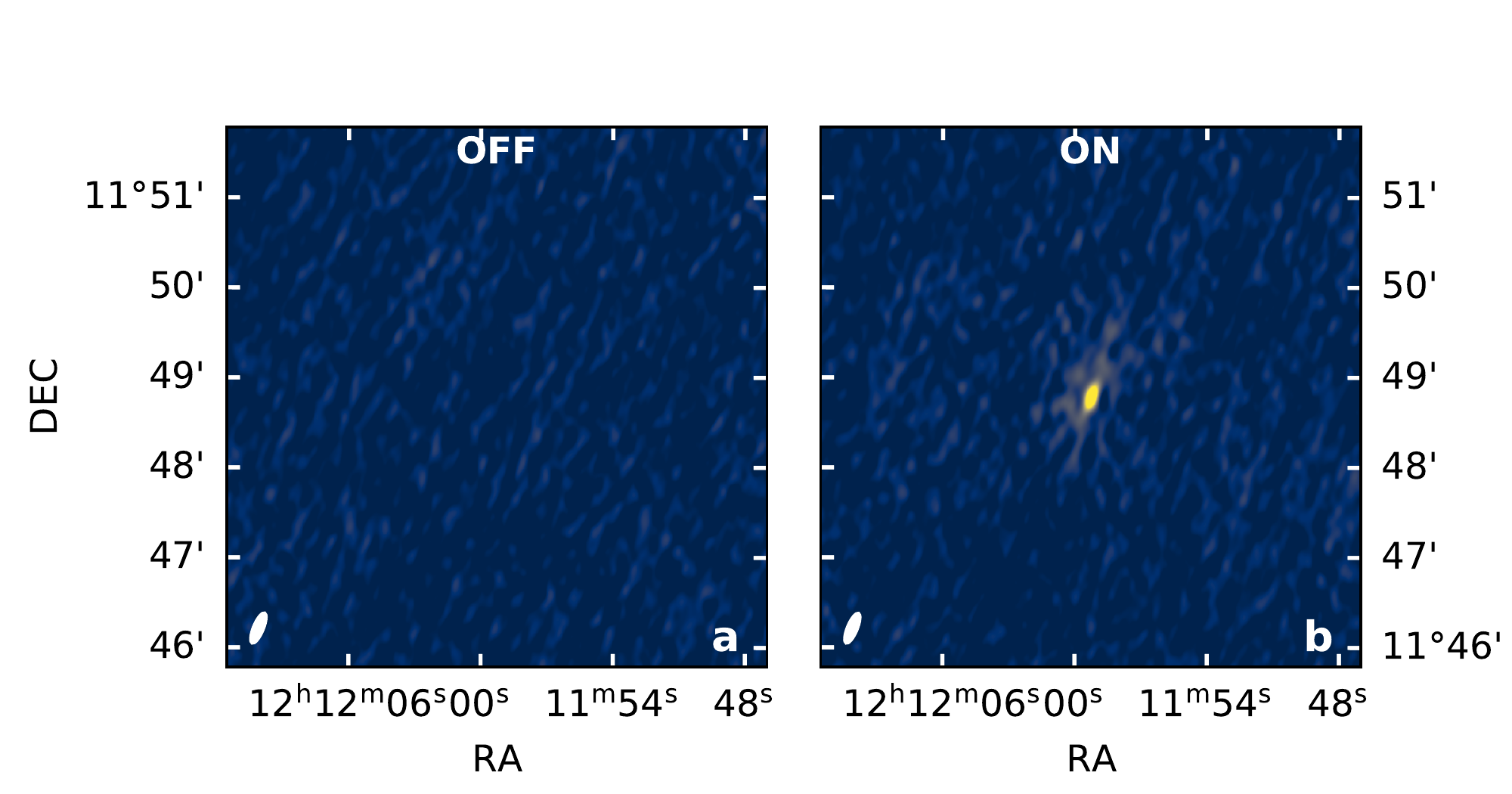}
    \caption{\textbf{Localisation of FRB~20240304B in the MeerKAT radio images.} Panel a (left) shows the OFF image before the burst, and Panel b (right) the ON image during the burst. The white ellipse at the bottom left corner shows the beam size.}
\label{fig:on_off}
\end{figure}

To obtain an initial estimate of the FRB location, we first generated dirty images for 11 time intervals around the central time of the 300\,ms of TB data. We identified a single source appearing only in the central time intervals, within the coordinates of the coherent beam number 689 where it was detected by the real time system. We then selected the time intervals where the source appeared to create a new ``ON" image with improved cleaning parameters. Despite strict RFI flagging of the frequency subbands, the FRB remains clearly visible in the ``ON" image. An ``OFF" image was also created by integrating the same amount of time but from a period where the FRB was absent. Finally, we generated a deep image by integrating the full 300\,ms of TB data (``FullInt"), which we used to perform the astrometric correction. All three images were created with 8192 pixels per side, and a pixel size of $1\arcsec$. The ON and OFF images are shown in Figure \ref{fig:on_off}.

Next, we identified the sources within the field-of-view (FoV) of the ON, OFF, and FullInt images using \texttt{PyBDSF} \cite{mohan_pybdsf_2015} with typical settings, and applied an astrometric correction.
To do this, we first identified radio source catalogues containing a sufficient number of sources within the FoV. The second epoch of the Very Large Array Sky Survey (VLASS 2.1) \cite{lacy_karl_2020} covers the whole sky above declinations $>-40\deg$, and it has a typical positional accuracy of 0.2\arcsec. We thus selected VLASS as a reference catalogue for the astrometric correction. 

We selected VLASS sources around the MeerKAT phase center that were identified as single Gaussians by \texttt{PyBDSF}, indicating they are unresolved. 
Next, we selected unresolved FullInt MeerKAT sources within the L-band half-power beam width ($1.1\deg$ in diameter) \cite{de_villiers_meerkat_2023}, and matched them to the VLASS positions using a 5\arcsec\ tolerance.
We computed a transformation matrix to adjust the MeerKAT source positions via an affine transformation, using the matched VLASS and uncorrected MeerKAT sources. The transformation matrix was derived from the weighted positional offsets between the reference (VLASS) and measured (MeerKAT) sources, where the weights were defined as the inverse of the quadratic sum of the VLASS and MeerKAT positional uncertainties, including a systematic 0.2\arcsec\ error on the VLASS positions. We performed this minimization using the least squares method from \texttt{SciPy}.
After applying the final transformation matrix to the sources in both the FullInt and ON images, we determined the astrometrically corrected position of the FRB to be RAJ$=$12:11:59.29 and DECJ$=+$11:48:46.86 (J2000 reference frame).

We computed the errors in RA and DEC after the transformation as the weighted mean separation in RA and DEC between the VLASS and MeerKAT corrected sources, which resulted in (0.17\arcsec, 0.35\arcsec) in RA and DEC respectively. 
The final FRB positional uncertainty was computed as the quadratic sum of the transformation errors, the \texttt{PyBDSF} source errors (0.09\arcsec, 0.25\arcsec), and the 0.2\arcsec\ VLASS systematic errors. This resulted in a final RA, DEC uncertainty on the FRB position of (0.28\arcsec, 0.48\arcsec).

\subsection{Ground-based optical and Infrared imaging}

We performed ground-based infrared imaging of the \frb\ field with the MMT and Magellan Infrared Spectrograph (MMIRS) on the 6.5-m MMT on 2024 May 27 UTC (Program UAO-G200-24A, PI Nugent). We obtained $J-$band imaging with a total exposure time of $46.5$ minutes under average seeing conditions of $1.3$ arcsec. These data were reduced using the \texttt{POTPyRI}\footnote{\url{https://github.com/CIERA-Transients/POTPyRI}} imaging reduction pipeline. This pipeline performs flat-field, bias, and dark frame correction and applies a World Coordinate System calibrated to Gaia DR3 to the calibrated, stacked data. Lastly, we determined the image zeropoint by performing aperture photometry on sources within the image and calibrating against 2MASS.

\begin{figure}
\centering
  \includegraphics[width=0.95\textwidth]{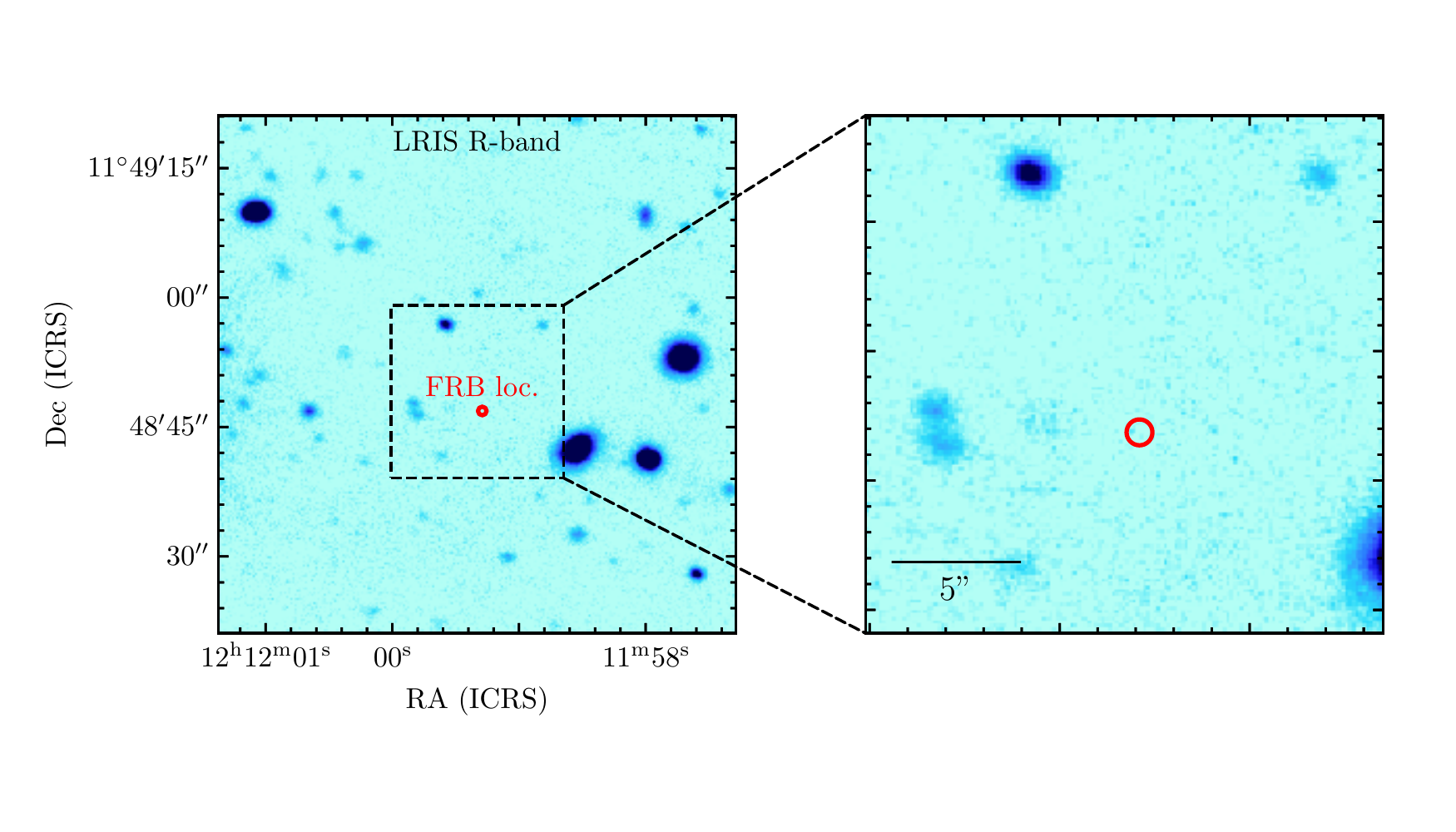}
    \caption{\textbf{Keck LRIS R-band image of the position of FRB~20240304B} {\it Left:} a $1\arcmin \times 1\arcmin$ region of the Keck LRIS R-band image centred around the radio localisation of the \frb{}. The image is complete to 26.8 mag. {\it Right:} a zoom-in ($20\arcsec \times 20\arcsec$) on the \frb{} localisation. The position of the \frb{} (and its uncertainty) is illustrated by the red circle in both panels. The \texttt{PATH} algorithm \cite{path} yielded host galaxy association posterior probabilities of $P\left( O|x\right) \sim 0$ for all identified galaxies in the vicinity of the FRB localisation meaning that the host galaxy is still unseen.}
\label{fig:lris_image}
\end{figure}

We obtained further optical imaging with the Low Resolution Imaging Spectrometer (LRIS) at the Keck I telescope on June 9th of 2024 (Program U299, PI Prochaska), under seeing conditions of about $0.7$ arcsec. The total exposure times were $55.2$ and $70.6$ minutes for the \textit{g} and \textit{R} filters, respectively, the d560 dichroic and $1\times1$ binning. The \textit{g} (\textit{R}) filter has an effective wavelength of $4730.52$ \AA\ ($6417$ \AA) and a full-width at half maximum (FWHM) of $1082$ \AA\ ($1185$ \AA)\footnote{\url{https://www2.keck.hawaii.edu/inst/lris/filters.html}}. 
The center of the pointing was offset from the FRB position toward the North by 30 arcsec to avoid the chip gap, and the time was split within eight pointing positions with cumulative offsets of (East, North): $(0,0)$, $(5,0)$, $(0,5)$, $(-5,0)$, $(-5,-10)$, $(5,0)$, $(-5,5)$, and $(0,5)$ arcsec. Similarly to the MMT observations, the data were reduced using the \texttt{POTPyRI} reduction pipeline. The resulting image is shown in Figure~\ref{fig:lris_image}.

\subsection{Disentangling the dispersion measure contributions}

The radio localisation of \frb{} places it 3.8 degrees from the centre of the Virgo Cluster, corresponding to an impact parameter of approximately 1.1 Mpc. To estimate the DM contribution from the intracluster medium (ICM) of Virgo, we follow the method outlined by \cite{alf+19}. Using Planck X-ray data \cite{planck2016}, we estimate the electron density out to twice the virial radius (2.4 Mpc) of the Virgo cluster as a function of radius. Our estimate of the electron density of $n_e = 7.6 \times 10^{-5}~\mathrm{cm}^{-3}$ is consistent with independent values published in the literature \cite{swa+17, erosita2024}, leading to an Virgo intracluster contribution of $\sim235\,\dmunits$. This is in line with the location of \frb{} in the outskirts of the Virgo cluster, where galaxy crowding is low. Further study of fluctuations in the electron density may offer insights into turbulence and small-scale structure within the ICM, and their impact on the total DM.

We further conducted ground-based integral field spectroscopic observations using KCWI on Keck II to identify any additional foreground structures along the line-of-sight that may contribute to the observed DM.
The total exposure times were $110$ and $50$ minutes with $2 \times 2$~binning in red and blue channels, respectively. We searched for the emission lines in the $20 \times 30\arcsec$ FoV around the FRB coordinates, and acquired spectra and redshifts of two galaxies. We estimated $z=0.31131$ for both of them based on the detected [OII] ($3726/3729{\rm \AA}$) and [OIII] ($4959/5007{\rm \AA}$) emission lines. On further investigation, we found a galaxy at the same redshift in the SDSS archival data, located within $\simeq 2\arcmin$ of the FRB position. We, therefore, confirm the presence of the galaxy group, along the FRB sightline. Additionally, upon inspection of the Legacy Survey DR9 catalogue we found another galaxy $\sim 25\arcsec$ away from FRB position with only photometric redshift $z_{\rm ph}=0.347\pm0.049$. Taking into account its small angular separation from the other galaxies, we included it into the group, adopting the measured group redshift. The \frb{} is located at the impact parameter $\sim240$~kpc relative to the estimated barycentre of this group. We estimate the lower limit on the galaxy group halo mass to be $\log_{10}\left( M_{\rm halo}/M_{\odot}\right) \simeq 12.8$. We adopt the modified NFW model to estimate the electron density distribution within the halo and the corresponding DM, yielding ${\rm DM_{group}} = 33\pm 13 \, \dmunits$ which is likely a lower limit. 

\subsection{Scattering analyses}

Potential sources of the scattering observed in FRBs are the galactic environments of the FRB host and the Milky Way, as well as the intervening galaxies and the diffuse gas in the IGM \cite{ocker+21, cordes+22}. The majority of scattering material is expected to be located in the ISM of the host and the Milky Way while, according to \cite{Macquart2013}, the diffuse intergalactic gas is an unlikely contributor to scattering due to its low density. Here it is also important to consider the relative impact of scattering from the Virgo cluster and the galaxy group intersected by the FRB sightline given their estimated contributions to the DM budget. 

\begin{figure}
\centering
  \includegraphics[width=0.6\textwidth]{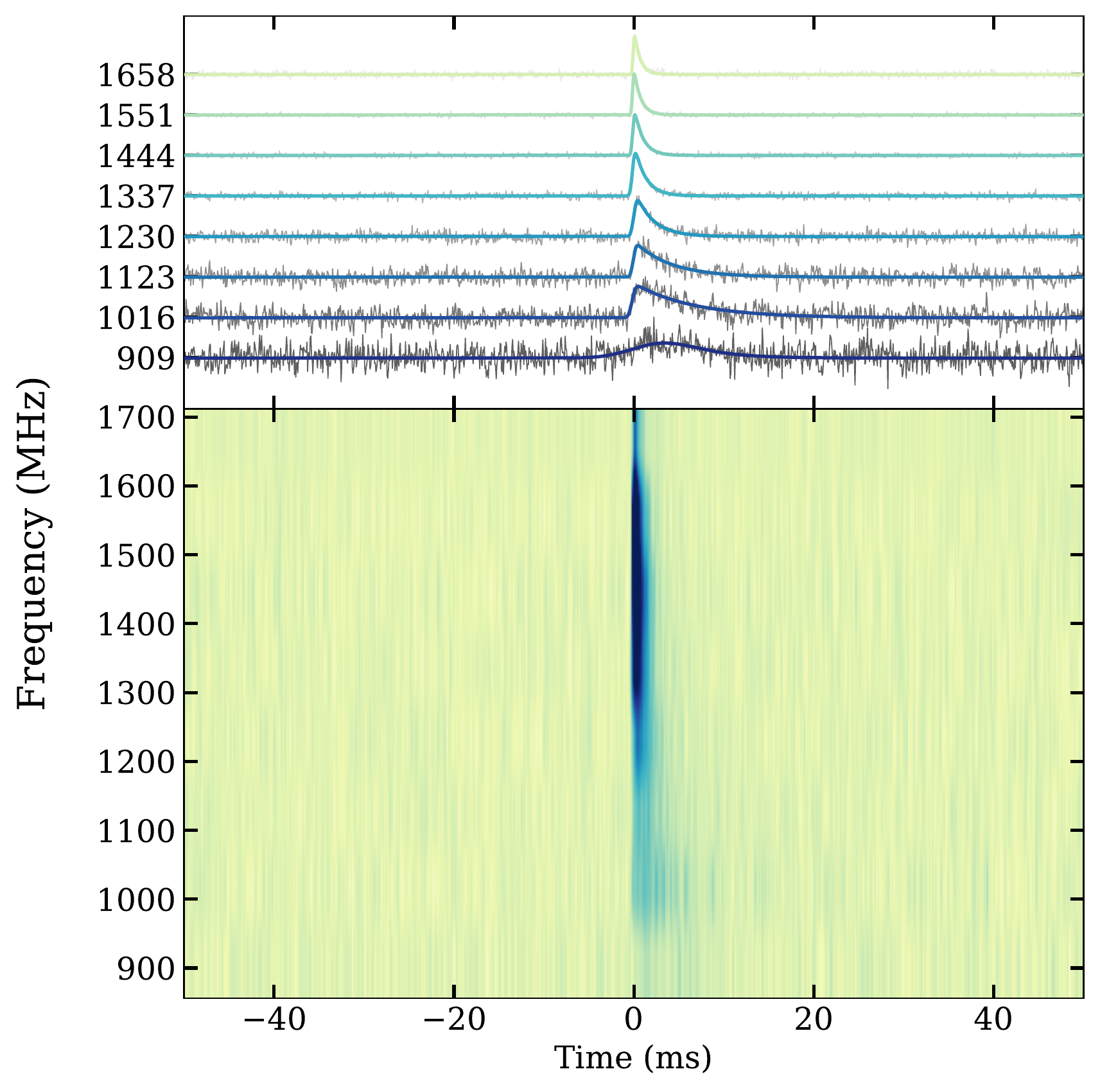}
  \includegraphics[width=0.7\textwidth]{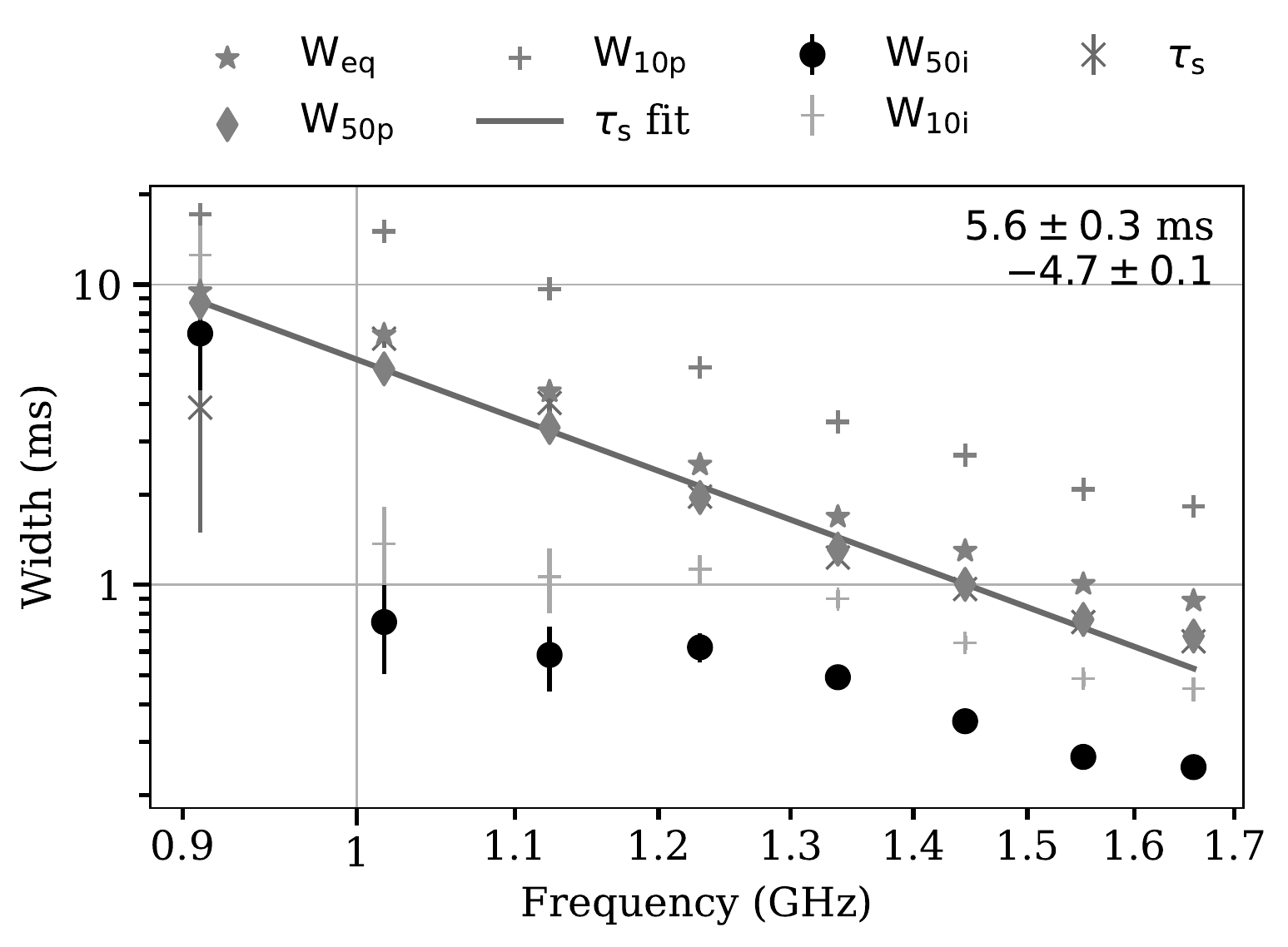} \\
    \caption{\textbf{Scattering analysis of FRB~20240304B.} \textit{Top panel:} Scattering fits to 8 subbands of the FRB data, each well fit by a single exponentially modified Gaussian.  \textit{Bottom panel:} We show the post-scattering pulse widths $W_\mathrm{50p}$ and $W_\mathrm{10p}$, the Gaussian intrinsic widths  $W_\mathrm{50i}$ and  $W_\mathrm{10i}$, the boxcar equivalent burst width $W_\mathrm{eq}$, and the scattering times $\tau_\mathrm{s}$ as a function of frequency. The best-fitting power law to the $\tau_\mathrm{s}$ data is displayed as a grey solid line. The text in the top right corner are the best-determined scattering time at 1~GHz and the scattering index.}
\label{fig:scatfit}
\end{figure}

We analysed the total intensity pulse profile of \frb{} obtained from the TB data by fitting a scattering model to the frequency sub-banded data (eight frequency sub-bands). The data were coherently dedispersed to a fiducial close-to-optimal DM before fitting. The model consisted of an `exponentially modified Gaussian', i.e.\ a Gaussian pulse profile convolved with a single-sided exponential pulse broadening function, characteristic of scattering in the thin-screen regime of turbulent ionised media \cite{ne2001}. The profile fits were performed using \textsc{scatfit}\footnote{\url{https://github.com/fjankowsk/scatfit}}, a custom Python-based software \cite{jankowski22} in version 0.3.2. \textsc{scatfit} utilises the Levenberg–Marquardt minimization algorithm, as implemented in \textsc{lmfit} \cite{newville2016}, to obtain an initial fit. It then explores the full posterior distribution of the fit parameters using the \textsc{emcee} Markov Chain Monte Carlo sampler \cite{Foreman-Mackey+2013}. For more details on the fitting technique, see \cite{jbc+23}. We iteratively ran the fits and refined the DM. The final, best-determined scattering-corrected DM is $2458.20\pm0.01\,\dmunits$. The DM uncertainty was determined by combining in quadrature the half-width at which the S/N versus trial DM curve decreased by unity (systematic uncertainty) with the statistical DM uncertainty from the scattering fit. The best-fitting scattering time at 1~GHz is $5.6 \pm 0.3~\text{ms}$, and the scattering index is $-4.7\pm 0.1$, shown as the solid line in Figure \ref{fig:scatfit}.

\subsection{JWST data acquisition  and data reduction}\label{sec:jwst}

The JWST observations utilized in this analysis were acquired as part of the JWST DD programme 6779 (PI Caleb). With no FRB host galaxy candidates available from ground-based imaging, the programme was designed to first obtain NIRCam imaging of the field to identify potential host galaxies, and then to follow up candidate hosts with the JWST/NIRSpec integral field spectrograph (IFS).

NIRCam permits simultaneous observations in its blue and red channels; hence, we opted for concurrent imaging in the F200W and F322W2 bands. We selected the F200W band because, based on the dispersion measure, we expected the galaxy to lie at $z \gtrsim 2$ and therefore display strong optical emission lines (e.g. \Halpha) in this band. In the red channel, we chose the F322W2 band to maximise the wavelength coverage of the source, thereby increasing the likelihood of detecting its continuum even if the galaxy was faint. Each band was observed with 9 groups per integration using the MEDIUM8 readout pattern, with 8 dithers implemented via the INTRAMODULEX primary dither pattern (comprising 4 primary dithers, each with 2 small grid dithers) to mitigate bad pixel effects, resulting in $\sim2$ hours of exposure time. The FRB was positioned on the Module B detector, which offer a higher throughput than those in Module A.

NIRCam images were obtained on 26 December, and the final reduced, combined images utilising the STScI JWST calibration pipleine revealed a potential FRB host candidate at RA=12:11:59.29 and Dec=+11:48:46.63. The candidate galaxy was measured to have an AB magnitude of $27.82\pm0.06$ in the F200W band and $28.73\pm0.03$ in the F322W2 band. Based on the NIRCam observations, we updated the JWST NIRSpec IFS location to centre near the host.

The NIRSpec IFS observations were obtained on 22 January using no filter (CLEAR) in the PRISM mode.  We chose this mode to maximise the wavelength coverage ($0.6\,\mu\mathrm{m}$--$5.3\,\mu\mathrm{m}$), thereby increasing the likelihood of determining the galaxy's redshift. The drawback is that PRISM/CLEAR mode has the lowest resolution (ranging between $R\sim50$ and $R\sim500$ \cite{Nanayakkara2024}), which necessitates higher line equivalent widths to secure a robust spectroscopic confirmation of the galaxy. Each exposure comprised 20 groups using the NRSIRS2 readout pattern. A 4-point dither was employed to mitigate bad pixel issues. In total, 16 exposures were obtained, resulting in an on-target exposure time of $\sim6.5$ hours.

NIRSpec raw data were reduced using the STScI calibration pipeline. The final combined cube was background‐subtracted by removing the median along the y‐axis of the NIRSpec detector. Analysis of the final cube with SAOImage DS9 revealed the presence of \Hbeta+\OIIId\ and \NIId+\Halpha\ (henceforth \Halpha) emission lines. To secure a robust redshift measurement, we extracted an optimally weighted one-dimensional spectrum \cite{Horne1986} of the source using {\tt mpdaf}\footnote{\url{https://mpdaf.readthedocs.io}}, weighted by the segmentation from {\tt Source Extractor} \cite{Bertin1996} applied to the \Halpha\ narrow-band image. To better constrain the uncertainties in the JWST spectrum, we employed a bootstrap resampling technique during Stage 3 of the pipeline. Specifically, we randomly resampled the 16 NIRSpec exposures with replacement, reducing the data 100 times. We then followed the same procedure as described above for background subtraction, source detection, and extraction. The per-pixel standard deviation of the 100 extracted spectra was adopted as the error for each pixel.

Data presented in this paper were reduced using the most up-to-date version of the JWST STScI public pipeline (v1.17.1) and the corresponding calibration files (CRDS context \texttt{jwst\_1322.pmap}) available at the time.

\subsection{Host Association using PATH}\label{sec:host_galaxy}

\begin{figure}
\centering
  \includegraphics[width=5.5 in]{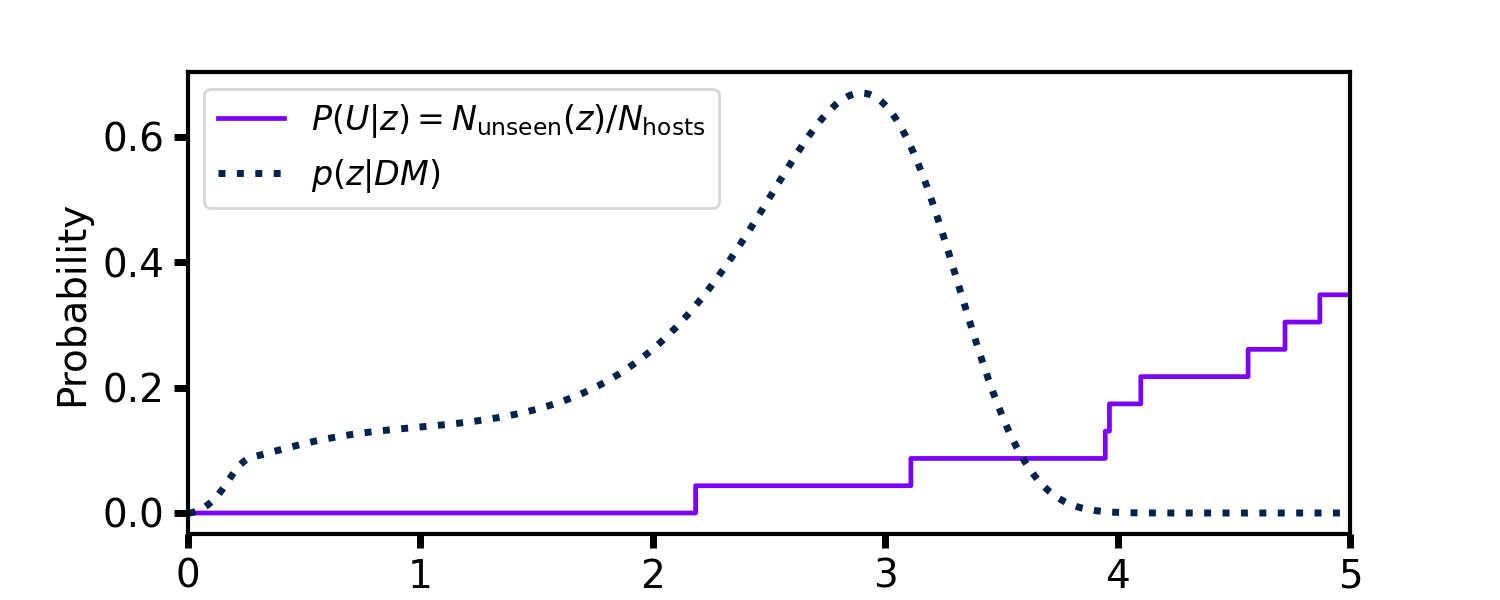}
    \caption{\textbf{Estimation of the unseen prior for PATH analysis.} The solid line traces the fraction of the 23 hosts in \cite{mrj+23} that would be fainter than the magnitude limit of our F200W band at the given redshift $N_{unseen}(z)/N_\mathrm{hosts}$; we use this as a coarse probability that a new FRB host at redshift $z$ would be unseen, or \puz. The dotted line is the best-fitting \pzdm\, probability density function for \frb{}.}
\label{fig:unseen-prior}
\end{figure}

As shown in \ref{sec:localisation}, the final localisation of \frb{} is RAJ$=$12:11:59.29$\pm 0.28\arcsec$ and DECJ$=+$11:48:46.86$\pm 0.48\arcsec$. We utilize the NIRCAM imaging in the F200W band to perform a PATH analysis \cite{path}. We assume the following priors for the PATH analysis. 
We estimate the unseen prior, \pu\,, using a coarse-grained approach based on FRB host luminosities, as described in Section B2.4 of \cite{mrj+23}. Our analysis considers the 23 known FRB hosts reported in \cite{mrj+23}, for which we compute the expected observational magnitudes if placed at plausible redshifts within the range $0 \leq z \leq 5$. We assume that ground-based K$_{s}$ magnitudes are analogous to those in the JWST F200W band. At any given redshift, the probability of a host being undetectable is given by \puz = $P(m>m_\mathrm{limit}|z) = N_\mathrm{unseen}/N_\mathrm{hosts}$ (Figure \ref{fig:unseen-prior}), which represents the fraction of hosts with magnitudes exceeding the K$_{s}$ band detection limit of 28.5. The overall unseen prior is then computed as $\pu\, = \int{ \puz \pzdm}dz = 0.035$, which we round to 5\%.

We obtain the candidate list for the priors on candidate galaxies by identifying the sources in the image. We use \texttt{Photutils} segmentation and find 340 candidates within 30 arcseconds of the FRB localization. We assumed inverse surface density priors ($P(O_i) \propto \frac{1}{\Sigma(m_i)}
$) with size weighting. We used galaxy number counts measured in the F200W filter. These counts were taken from \cite{Windhorst_2023}, which provides galaxy counts per unit area per 0.5 mag as a function of AB magnitude in this filter. This replaces the r-band-based counts used in \cite{path}, ensuring consistency with our observational data. We model the angular offset distribution of the transient relative to the candidate galaxy centre using an exponential prior, characterized by a scale length equal to $50\%$ of the candidate galaxy's half-light radius. We truncate the distribution at a maximum offset of 6 times the half-light radius. 

We find the posterior ($\pox$) for the top candidate to be \pathpost\ and the posterior for the unseen host (\pux) to be \puxpost. The priors and posteriors for the top 5 candidates are shown in Table \ref{tab:path_candidates}.

\subsection{Redshift measurements of host}\label{sec:slinefit}

We employed \slinefit\footnote{\url{https://github.com/cschreib/slinefit/tree/master}} \cite{Schreiber2018a} to derive a spectroscopic redshift. We fitted the \Halpha, \Hbeta, \OIIId, and \SII\ emission lines using single Gaussian profiles. Due to the low resolution of the PRISM data, \NII\ and \Halpha\ cannot be resolved separately; however, given the lower metallicity inferred for our source (see Section~\ref{sec:sed_fitting}), we assume that the contribution from \NII\ is negligible \cite{Pettini2004}. As our segmentation encompasses the full galaxy, and to account for potential strong velocities and outflows, we allowed the line widths to vary freely between 500 and 5000 km/s when fitting the integrated one-dimensional spectrum.

Using \slinefit, we obtained a best-fit redshift of $z_{\mathrm{spec}} = 2.148 \pm 0.001$. Redshift uncertainties were estimated from 200 Monte Carlo iterations, and a global residual scaling factor of 1.1 indicates that our bootstrapping technique accurately constrained the uncertainties in the JWST NIRSpec observations.

\subsection{Stellar Population Analysis in the host}\label{sec:sed_fitting}

We utilised spectral energy distribution fitting codes, \fastpp\ \cite{Kriek2009,Schreiber2018a} and \prospector\ \cite{Johnson2021a}, to derive the stellar population properties of the host galaxy.

First, we employed \fastpp\ with a delayed exponentially declining star formation history (SFH), defined as

\begin{equation} 
\mathrm{SFR}(t) = t\,\exp\left(-\frac{t}{\tau}\right).
\end{equation}

\noindent where $t$ represents time and $\tau$ is the characteristic timescale.  We adopted the \cite{Bruzual2003} stellar population models, a \cite{Calzetti2000} dust law, and a \cite{Chabrier2003} initial mass function (IMF), allowing the stellar metallicity to vary between $Z=0.004$ and $0.05$, the full range of models available in \cite{Bruzual2003} . As these models do not account for nebular emission line contributions, emission lines were masked in the observed spectrum prior to fitting. The observed spectrum was convolved with the instrument's line spread function. 

\fastpp\ yielded a best-fit stellar mass of $\log (M_*/M_\odot) = 6.89^{6.95}_{6.88}$ and a stellar metallicity of $Z=0.004$ (20\%\,\zsol)  for the FRB host. \fastpp\ fitting indicates the galaxy to have a negligible amount of dust ($A_V = 0.0^{0.05}_{0.00}$) and a very young age with 50\% of its stellar mass only formed within the last $<10$ million years.

Next, we used \prospector\ with a binned SFH parametrisation to obtain stellar population properties of the source. Following \cite{Leja2019a} we used seven time bins to parametrise the SFH. In terms of lookback time, the time bins for each galaxy were defined as follows. The first two bins were fixed at 0–30~Myr and 30–100~Myr, respectively, while the final bin spanned from 85\% to 100\% of the Universe's age. The remaining four bins were evenly distributed in logarithmic intervals between 100~Myr and 85\% of the Universe's age. 
The SFR was allowed to vary freely across these seven time bins ($j$), parametrised by six logarithmic ratios, $\log_{10}(\mathrm{SFR}_j/\mathrm{SFR}_{j+1})$.We allowed the stellar metallicity, dust optical depth, and stellar mass to vary freely. The stellar metallicity was permitted to range between $-0.5 < \log_{10}{Z/Z_\odot} < 2.5$, the dust optical depth (following the \cite{Calzetti2000} dust law) between $0 < \tau_\mathrm{dust} < 4.0$, and the stellar mass within $6.0 < \log_{10}(M_*/M_\odot) < 12.0$. 
We used a  \cite{Chabrier2003} IMF and with input stellar population models generated using the C3K \cite{Conroy2009} stellar library. To account for contribution from the observed nebular emission lines, we marginalised over a {\tt cloudy} \cite{Ferland2017} input grid with variable ionisation parameter between $-4.0 < U < -1.0$. We fixed the gas-phase metallicity  to match the stellar metallicity. Finally, each emission line was fitted with a single-component Gaussian and convolved with the instrument's line spread function.

\prospector\ provided us with a maximum a posteriori observed stellar mass of $\log (M_*/M_\odot) = 6.984^{7.109}_{6.979}$,  a stellar metallicity $Z/Z_\odot=0.109^{0.112}_{0.104}$, and negligible amount of dust. The ionisation parameter was U=$-1.0$, suggesting highly ionising conditions in this galaxy. The SFH reconstruction confirmed the young age of the galaxy with it reaching 50\% of formed stellar mass in the last 30-100~Myr time bin. The SFR average over the last 100~Myr is found to be $0.1$\,\msol/yr. 

Based on analysis using \fastpp\ and \prospector, we expect the FRB host to be a relatively low mass galaxy, with low metallicity, negligible amount of dust, and undergoing intense star formation with a sSFR$\sim 10$/Gyr.


\section{Supplementary Text}
The Supplementary Text section can only be used to directly support statements made in the main text
e.g. to present more detailed justifications of assumptions, investigate alternative scenarios,
provide extended acknowledgements etc.
Material in this section cannot claim results or conclusions that weren't mentioned in the main text.
To refer to this section from the main text, just write (Supplementary Text).

\subsection{Interpreting the scattering}

The expected Milky Way contribution to the observed scattering timescale of 5.6~ms at 1 GHz is approximately $\sim6\,\mu$s based on the NE2001 Galactic electron density model \cite{ne2001}. If we assume that the scattering originated in a scattering screen located in the host galaxy, factoring in the redshift dependence to the scattering wavelength $(1+z)$, and the scattering timescale $(1+z)^{-4}$, we estimate a scattering timescale of 179 ms at 1 GHz in the host galaxy reference frame. 
For Galactic sources, which are embedded in the scattering medium, wave sphericity causes less pulse broadening than for plane waves propagating through the same medium. Scattered waves from a distant source are effectively planar when they reach the Galaxy. This implies that scattering in the host should be a factor of three larger for the same scattering strength \cite{cordes+16}.
Decreasing the pulse broadening time of the FRB by a factor of three to account for the plane-wave scattering geometry, we estimate a scattering timescale of $\sim60$~ms in the host, which we consider to be a plausible value indicating that the host may play a 
major role for scattering. 

For the case of the galaxy group at $z = 0.31131$, Equation 47 in \cite{Macquart2013} indicates that to obtain a scattering time on the order of milliseconds requires a scattering measure (SM) of $\mathrm{SM} \sim 10^{15}\,{\rm m^{-17/3}}$, which is consistent with the values expected for galaxies in that work. If scattering in the group is driven by refraction, the formalism in (Equations 20 and 21 in \cite{Mas-Ribas2025}) requires that the gas clouds inducing scattering be smaller than $\sim0.1$ pc, and to have electron densities of $n_{\rm e} = 0.5\,{\rm cm^{-3} (r_{\rm c}/0.1\,{\rm pc})}$. 
This implies that the gas clouds should be about a hundred times below the minimum size 
dictated by the so-called scattering scale \cite{McCourt2018}, suggesting that diffractive scattering may be a more likely scenario instead. Furthermore, the geometrical term characterizing the relative distances between the source, the scattering screen and the observer, is a factor of $\approx100$ smaller for the Virgo cluster than for the galaxy group at $z=0.31131$, owing to the proximity of the cluster to the Milky Way. This rules out the contribution from the cluster in comparison to that from the  galaxy group. 

Overall, both the host galaxy and the galaxy group may substantially contribute to the observed scattering. 
Owing to the $(1+z)^2$ dependence of the scattering measure, however, the required SM value for the host should be almost an order of magnitude higher than that of the group. This may be justified by considering the path through the ISM for the case of the host,  while the sightlines through the galaxies in the group traverse the less dense circumgalactic media. Further data and analyses are required to set tighter constraints on the individual scattering contributors.

\subsection{Rotation Measure analyses}

\begin{figure}
\centering
  \includegraphics[width=0.9\textwidth]{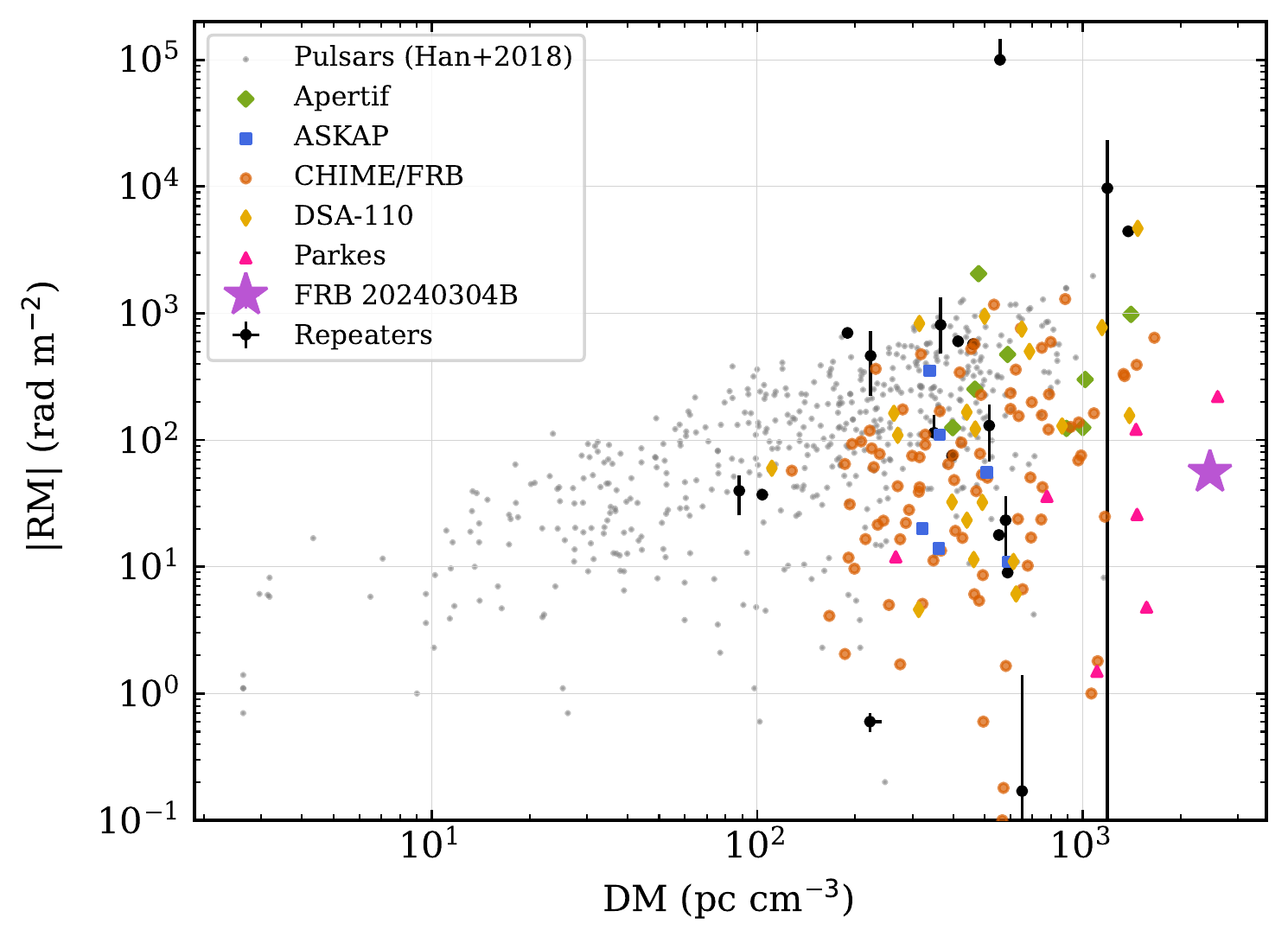}
    \caption{\textbf{Observed RMs (absolute value) of both one-off and repeating FRBs are shown as a function of observed DM, alongside Galactic pulsars for comparison.} The $|\mathrm{RM}|$ of \frb\ is shown as a purple star, grey dots show pulsar RMs \cite{han+2018}, green circles represent RMs measured with Apertif \cite{pastor-marazuela_comprehensive_2025}, blue squares are one-off FRBs detected by ASKAP, orange circles are from CHIME/FRB \cite{pandhi_polarization_2024}, yellow diamonds from DSA-110 \cite{sherman_deep_2024}, and pink triangles from Parkes. The mean RMs of repeating FRBs are shown as black circles, with error bars indicating the observed RM range for each source.}
\label{fig:DMvsRM}
\end{figure}

Using the saved channelized complex voltage data for the burst, we were able to study the polarimetric properties of the burst. We formed a coherent beam pointing towards the best determined position of the FRB from the interferometric localisation. The coherent beam contains information at the highest time resolution and with all the Stokes Parameters. Then we ran the \texttt{rmsynth} software as part of the \textsc{PSRSALSA} software suite for data analysis of time-domain radio data for pulsars~\cite{psrsalsa}. The software uses the Fourier relationship between the Faraday Rotation and the linear polarization of the incoming radio wave to produce the Rotation Measure Spread Function (RMSF). This RMSF can then be used to find peaks in the Fourier space to obtain the rotation measure of the burst. Using this technique, we obtained a rotation measure of $-55.6\pm0.5$\rmunits. Assuming the observed RM is local to the source, its rest-frame value increases by a factor of $(1+z)^2$, giving an intrinsic RM of $\sim -550$\,\rmunits. This places it close to the median for most one-off FRBs, many of which show minimal Faraday rotation (see Figure \ref{fig:DMvsRM}) \cite{caleb+2018, mckinven+23, sherman+23, mpp+23, pastor-marazuela_comprehensive_2025}. However, some one-off FRBs also exhibit extreme RM values which could reach absolute RM values exceeding 1000\,\rmunits in the source reference frame (see Figure \ref{fig:DMvsRM}) \cite{pastor-marazuela_comprehensive_2025}.
The high rest-frame RM of \frb{} suggests that the source is embedded in a magnetized circumburst environment, such as a star-forming region or remnant of a recent supernova, indicating that some high-redshift FRBs originate in dense, dynamic and magneto-ionic environments \cite{pastor-marazuela_comprehensive_2025}.
The burst shows strong linear polarization ($L/I = 49\%$) and marginal circular polarization ($|V|/I = 3\%$) with no strong evidence for polarization position angle changing across the pulse. We do note that the polarization fraction estimate is not corrected for any instrumental effects due to the offset from the boresight of the telescope due to the unavailability of the Jones matrices at the position of the FRB. We account for these errors by adding an additional 2$\%$ uncertainty on the obtained values. 

It is well established from the observations of pulsars that scattering does not only affect the total power of the shape of the total intensity pulse profile, but also its measured polarization properties (e.g., \cite{LiHan2003, kj2008, aris2009}). While all Stokes parameters are affected, including potentially the inferred degree of polarization, the impact of scattering is easily recognisable from a flattening of the position angle (PA) swing towards latter pulse phases \cite{LiHan2003}. In the presence of scattering, similar effects should also be expected for FRBs. 
The PA of \frb{}\ is predominantly flat and so determining whether any of that flatness is intrinsic is difficult given the presence of scattering.

The measured RM is small, suggesting opposing contributions from different components of the IGM along the line of sight. This can happen due to structured magnetic fields in opposite directions to the line-of-sight. Previous studies of this line-of-sight, particularly the Virgo cluster has revealed ordered magnetic field ~\cite{vallee1991}. This might suggest that the FRB encounters a constant magnetic field in the Virgo cluster that is in the opposite direction compared to the general direction of the magnetic field in the IGM, resulting in a small value of the RM.

\subsection{Energetics}

To determine the flux and fluence of \frb, we followed the method described in \cite{2023Jankowski}, and applied the radiometer equation \cite{dewey_search_1985} adapted for single-pulses;

\begin{equation} \label{eq:radiometer}
    S_{\text{peak}} = \text{S/N} \beta \eta_{\text{b}} \dfrac{T_{\text{sys}} + T_{\text{sky}}}{G \sqrt{b_{\text{eff}} N_{\text{p}} W_{\text{eq}}}} a_{\text{IB}}^{-1},
\end{equation}

\noindent where $S_{\text{peak}}$ is the peak flux density, $\beta$ is the digitisation loss factor, $\eta_{\text{b}}$ the beamforming efficiency, $T_{\text{sys}}$ and $T_{\text{sky}}$ the system and sky temperatures respectively, $G$ is the telescope gain, $b_{\text{eff}}$ the effective bandwidth in Hz, $N_{\text{p}}$ the number of polarisations, and $W$ is the pulse width in seconds. $a_{\text{IB}}$ is the attenuation factor of the IB, which depends on the observing bandwidth and the angular separation from the boresight 
\footnote{To compute attenuation factors: \url{https://github.com/BezuidenhoutMC/beam-corrections}} \cite{de_villiers_meerkat_2023}. 

The parameters relating to the telescope performance are the gain $G\sim2.77$\,K\,Jy$^{-1}$ for 64 antennas \cite{bailes_meerkat_2020}, scaled down to the 62 antennas that were being used when the FRB was detected, $N_\text{p}=2$, $T_{\text{sys}}=19$\,K, and the digitisation loss factor and beamforming efficiency are both close to unity. 
The FRB had an TB S/N=114.7 and a scattering timescale of 5.6\,ms at 1~GHz. 
To estimate the sky temperature, we use the latest version of the Haslam 408\,MHz sky model \cite{remazeilles_improved_2015}, scaled to the L-band central frequency of 1284\,MHz using the python module \texttt{PyGDSM} \cite{price_pygdsm_2016}, resulting in $T_\text{sky}=3.54$\,K at the FRB coordinates.
After removing the frequency channels contaminated by RFI, we estimate an effective bandwidth $b_\text{eff}=614.44$\,MHz.
We find the attenuation factor to be $a_\text{IB}=0.75$.

By substituting the values above in Eq.~\ref{eq:radiometer}, and assuming the error to be given by an uncertainty of one unit in the S/N, we find the flux to be $S_\text{peak}=0.49\pm0.01$\,Jy. 
We convert this to a fluence $F_{\nu}$ by multiplying by the width of the burst in milliseconds, and we find $F_{\nu}=2.75\pm0.05$\,Jy\,ms.

The isotropic equivalent energy density of transients is related to the observed fluence \cite{cordeschatterjee2019} via

\begin{equation} 
E_{\nu} = \frac{4 \pi D_{L}^2\Delta\nu}{(1+z)} F_{\nu}.
\end{equation}

\noindent For a redshift of \redshift{} and a luminosity distance of $D_{L} = 17.4$~Gpc, we estimate an energy density of $E_{\nu} = 2 \times 10^{41}$~erg over a bandwidth, $\Delta\nu$ of 614.44~MHz.





\begin{figure}
\centering
  \includegraphics[width=0.9\textwidth]{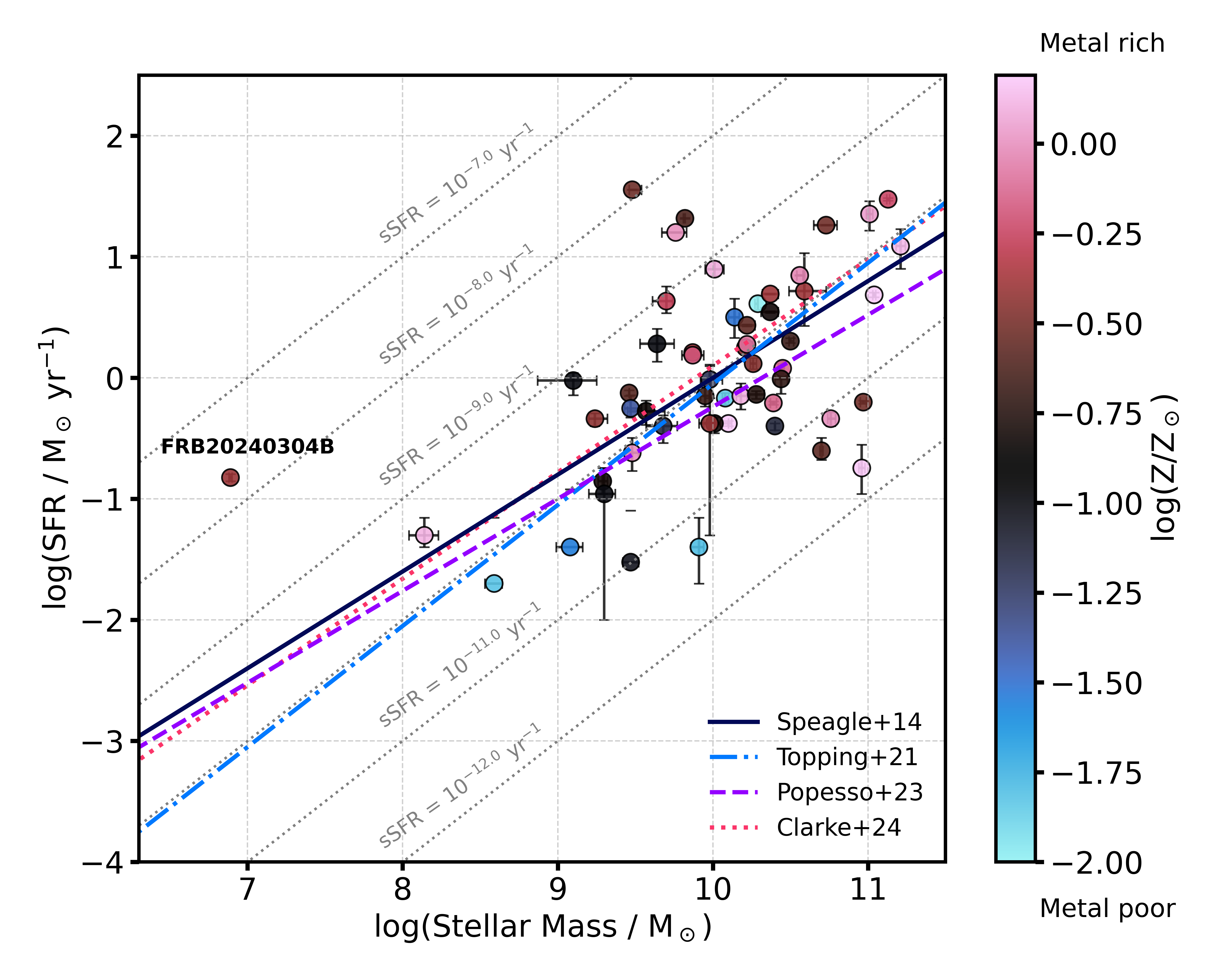}
    \caption{\textbf{Star formation rate vs. stellar mass for the complete published FRB host galaxy sample.} Lines of constant sSFR are shown as dotted lines along with various SFMS lines. The colour scale indicates the metallicity, with galaxies having log(Z/Z$_\odot$) $< -1$ considered metal poor and those with log(Z/Z$_\odot$) $> 0$ classified as metal rich.}
\label{fig:sfr_mass_metallicity}
\end{figure}


\begin{figure}
\centering
  \includegraphics[width=0.9\textwidth]{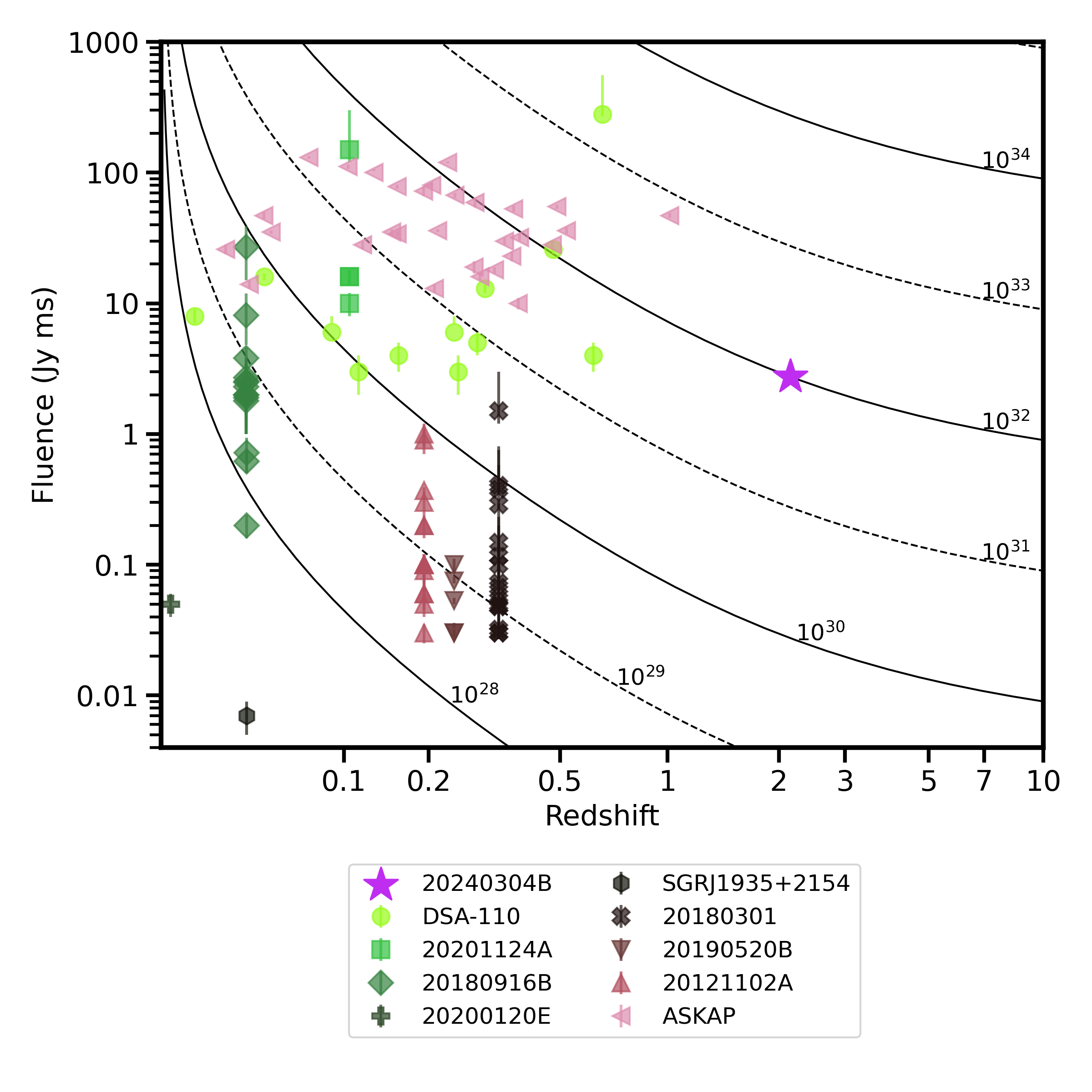}
    \caption{\textbf{Logarithmic plot showing fluence as a function of redshift for localized FRBs (adapted from \cite{shannon+18}).} Additional notable FRBs are labelled by their Transient Name Server designations, with the ``FRB" prefix removed. The energy density is indicated by the curved solid and dashed contours, measured in erg Hz$^{-1}$. The fluence of SGR~J1935$+$2154 is plotted at the redshift of the host galaxy of the nearest FRB, FRB~20180916B.}
\label{fig:shannonplot}
\end{figure}


\begin{table}
\caption{\textbf{PATH Results for the top 5 candidates.}}
\label{tab:path_candidates}
\begin{tabular}{cccccccc}
\hline
RA (J2000) & Dec (J2000) & Half-light radius  & F200W & Separation  & $P(O_i)$ & $P(x|O)$ & $P(O|x)$ \\
Degrees & Degrees & arc-sec & AB[mag] & arc-sec &  &  &  \\
\hline
182.99704 & 11.81295 & 0.10291 & 28.05016 & 0.23454 & 0.00055 & 1.00574 & 0.97518 \\
182.99695 & 11.81252 & 0.03059 & 30.15953 & 1.81018 & 0.00019 & 0.00066 & 0.00022 \\
182.99695 & 11.81360 & 0.13859 & 26.36513 & 2.13646 & 0.00130 & 0.00007 & 0.00016 \\
182.99770 & 11.81326 & 0.18794 & 25.79974 & 2.47192 & 0.00176 & 4.10e-10 & 1.27e-09 \\
182.99667 & 11.81385 & 0.11735 & 26.76791 & 3.26209 & 0.00106 & 1.82e-12 & 3.38e-12 \\
\hline
\end{tabular}

\end{table}







\end{document}